\documentclass[%
 reprint,
superscriptaddress,
 amsmath,amssymb,
 aps,
]{revtex4-1}

\usepackage[pdftex,hiresbb]{graphicx}% Include figure files
\usepackage{dcolumn}% Align table columns on decimal point
\usepackage{bm}% bold math
\usepackage{here}
\usepackage[version=3]{mhchem}

\renewcommand{\Vec}[1]{\mbox{\boldmath$#1$}}

\begin{document}

%\preprint{APS/123-QED}

\title{
Superconducting mechanism for the cuprate Ba$_2$CuO$_{3+\delta}$ based on a multiorbital Lieb lattice model}% Force line breaks with \\

\author{Kimihiro Yamazaki}
\affiliation{%
Department of Physics, Osaka University, Machikaneyama, Toyonaka, Osaka 560-0043, Japan
}%

\author{Masayuki Ochi}%
\affiliation{%
Department of Physics, Osaka University, Machikaneyama, Toyonaka, Osaka 560-0043, Japan
}%
\author{Daisuke Ogura}
\altaffiliation[Present affiliation: ]{Hitachi, Ltd, Marunouchi, Tokyo, 100-8280, Japan.}%Lines break automatically or can be forced with \\
\affiliation{%
Department of Physics, Osaka University, Machikaneyama, Toyonaka, Osaka 560-0043, Japan
}%

\author{Kazuhiko Kuroki}
\affiliation{%
Department of Physics, Osaka University, Machikaneyama, Toyonaka, Osaka 560-0043, Japan
}%

\author{\\Hiroshi Eisaki}
\affiliation{%
National Institute of Advanced Industrial Science and Technology (AIST), Tsukuba 305-8568, Japan
}%
\author{Shinichi Uchida}
\affiliation{%
National Institute of Advanced Industrial Science and Technology (AIST), Tsukuba 305-8568, Japan
}%
\affiliation{%
Institute of Physics, Chinese Academy of Science, Beijing 100190, China
}%
\author{Hideo Aoki}
\affiliation{%
National Institute of Advanced Industrial Science and Technology (AIST), Tsukuba 305-8568, Japan
}%
\affiliation{%
Department of Physics, University of Tokyo, Hongo, Tokyo 113-0033, Japan
}%



%\affiliation{
%\affiliation{
%\affiliation{%



\begin{abstract}
For the recently discovered cuprate superconductor $\mathrm{Ba_{2}CuO_{3+\delta}}$, we propose a lattice structure which resembles the model considered by Lieb to represent the vastly oxygen-deficient material. We first investigate the stability of the Lieb-lattice structure, and then construct a multiorbital Hubbard model based on first-principles calculation. By applying the 
fluctuation-exchange approximation to the model and solving the linearized Eliashberg equation, we show that $s-$wave and $d-$wave pairings closely compete with 
each other, and, more interestingly, that the intra-orbital and inter-orbital pairings coexist. We further show that, if the energy of the $d_{3z^2-r^2}$ band is raised to make it ``incipient'' with the lower edge of the band close to the Fermi level within a realistic band filling regime, $s\pm$-wave superconductivity is strongly enhanced. We reveal an intriguing relation between the Lieb model and the two-orbital model for the usual K$_2$NiF$_4$ structure where a close competition between $s-$ and $d-$wave pairings is 
known to occur.  The enhanced superconductivity in the present model 
is further shown to be related to an enhancement found previously in the bilayer Hubbard model with an incipient band.
\end{abstract}

\pacs{Valid PACS appear here}% PACS, the Physics and Astronomy
\maketitle

%\tableofcontents

\section{INTRODUCTION}

More than 30 years have passed since the discovery of the high-$T_c$ cuprates, but a full understanding of their physics remains one of the most challenging problems in the condensed matter physics~\cite{Review}.  
However, one strong consensus has been reached: The $\mathrm{CuO}_{2}$ planes play an 
essential role in the occurrence of superconductivity. 
Namely, the cuprates have a layered perovskite crystal structure, where a copper atom is surrounded by oxygens, 
typically with an octahedral coordination. Since the octahedron is elongated in the $c$-axis direction, the crystal field splitting makes 
the $3d_{x^{2}-y^{2}}$ orbital have the highest energy among the $3d$ orbitals. 
Hence, the $d^9$ electron configuration results in a situation where the electronic structure can be regarded as basically a single-band system. Indeed, some of the present authors have shown that there is a systematic material 
dependence, in which $T_c$ is basically increased as the one-band character 
($3d_{x^{2}-y^{2}}$) becomes stronger, i.e., when the energy of the $3d_{3z^{2}-r^{2}}$ orbital is lowered below that of $3d_{x^{2}-y^{2}}$, 
which is realized for higher  apical oxygen heights~\cite{cu1, cu2, cu3, cu4}. 

The recent experimental discovery by Li {\it et al.}~\cite{Ba} of 
another type of cuprate superconductor, $\mathrm{Ba}_{2}\mathrm{CuO}_{3+\delta}$, 
is remarkable in this context.  
The material, having a layered structure, exhibits $T_{c}=73\,\mathrm{K}$, which is much higher than that of ``214" $\mathrm{La}_{2-x}\mathrm{Sr}_{x}\mathrm{CuO}_{4}$~\cite{La} with $T_c\simeq 40\,\mathrm{K}$, but more interestingly, a dramatic 
feature, among others not seen in conventional cuprates, is that a large amount of oxygen deficiencies exist in the $\mathrm{CuO}_2$ planes~\cite{Odeficiencies}.  Details of the sample preparation is reported to be as follows:  
Ba$_2$CuO$_{3+\delta}$ samples  are synthesized in a
tetragonal symmetry at a much higher pressure (18\,GPa) than usual, 
and at a temperature of 1000$^\circ$C, in a polycrystalline form.  
This is in contrast with a lower-pressure synthesis in which 
only an orthorhombic phase is synthesized.  
This implies that the tetragonal phase, even if metastable, is 
stabilized with the high-pressure synthesis.  
The excess oxygens O$_\delta$ are also added in the Cu-O planes by the extremely high pressure synthesis with $\delta\simeq 0.2$.  
This immediately raises a puzzle regarding the origin of the high $T_c$ because the $\mathrm{CuO}_2$ planes should simply be disrupted at this level 
of O deficiency from the conventional Cu-O plane. 
Another notable feature in $\mathrm{Ba}_{2}\mathrm{CuO}_{3+\delta}$ 
is that the combination of the oxygen content of $3+\delta\simeq 3.2$ and the $+2$ valence of $\mathrm{Ba}$ should make the electron configuration 
significantly deviate from $d^{9}$, namely, an unprecedented amount of holes 
(as large as $\sim$40\%) exist.  This sharply contrasts 
with the conventional wisdom for the cuprates that superconductivity is 
optimized around 15\% hole doping~\cite{Review}. Yet another curious 
feature is that each CuO octahedron is compressed rather than elongated along the $c$ axis 
with the apical oxygen height smaller than the in-plane Cu-O distance, so that the Cu $3d_{3z^{2}-r^{2}}$ orbital should be higher approaching that of $3d_{x^{2}-y^{2}}$, and so  a multiband, multiorbital situation is expected. These features are all in strong contradiction with the high-$T_c$ condition for the conventional cuprates, which suggests that an alternative pairing mechanism may be at work in this new material. Indeed, a number of theoretical studies have proposed various pairing mechanisms based on various lattice structures~\cite{BaRPA,LiLiu,PhysRevMaterials.3.044802,LeJHu,NiZou,WangZhang}. The experimental finding of $\mathrm{Ba}_{2}\mathrm{CuO}_{3+\delta}$ may also shed a light on the previous finding of $\mathrm{Sr_{2}CuO_{3+\delta}}$~\cite{Sr213a, Sr213b, Sr213c}, which also possesses a large amount of oxygen deficiencies and a $T_c$ as high as $\simeq 90\,\mathrm{K}$ but a much lower superconducting fraction 
than in the $\mathrm{Ba_{2}CuO_{3+\delta}}$.

Given this background, a theoretical challenge is that 
how we can construct a model and fathom the structure of 
the gap function for the material, which has hugely oxygen-depleted \ce{CuO2} planes.   
Assuming that the deficiencies are ordered, some candidates for the crystal structure have been proposed. 
Liu {\it et al.} propose a chain-type structure, which actually exists in $\mathrm{Sr_{2}CuO_3}$~\cite{Srchain1,Srchain2,Srchain3,PhysRevMaterials.3.044802}. Li {\it et al.} predict a ladder-type lattice based on an automated structure inversion method~\cite{LiLiu}. Le {\it et al.} propose a structure where a matrix of $\mathrm{Ba}_{2}\mathrm{CuO}_4$ with CuO$_2$ planes is embedded in the background of $\mathrm{Ba}_{2}\mathrm{CuO}_3$~\cite{LeJHu}. More recently, another type of lattice dubbed as the brick-wall model has been proposed~\cite{WangZhang}. 

Thus, the lattice structures considered so far (other than the conventional $\mathrm{K}_{2}\mathrm{NiF}_{4}$ type) 
have one-dimensional natures in some sense or other, but   
an experiment~\cite{Ba} suggests the material has 
tetragonal symmetry.  
This has motivated us to propose here another structure, depicted in Figs.~1(a), and 1(c), as a candidate for the undoped $\mathrm{Ba_{2}CuO_3}$ (``213" composition), where by doping we mean adding excess oxygens.  
We call the proposed structure the ``Lieb-lattice type,'' since it resembles the model considered by Lieb~\cite{Lieb} if we focus on the Cu sites 1, 2, and 3 in Fig.~1, and ignore Cu site 4, which is shown to be electronically irrelevant.  The model considered by Lieb possesses a flat band in the band structure, and, 
in the context of magnetism, it is theoretically  proven that ferromagnetism 
occurs at half-filling when the on-site repulsive interaction $U$ is turned on. A superconducting mechanism exploiting the flat band of the Lieb lattice has also been proposed~\cite{LiebSC}. 
Lieb originally considered a class of models with different numbers of sublattice sites, and superconductivity in such a model in a quasi-1D structure has also been studied with the density-matrix renormalization group~\cite{KobayashiAoki}.
Here, however, we shall see that the model derived in the present study 
is actually {\it distinct} from the original (single-orbital) Lieb model, 
%in its original sense 
since the present material inherently has a multiorbital nature, as we shall show.

We start with an investigation of the stability 
of the Lieb lattice in terms of the total energy and phonon calculations 
for the lattice structure of $\mathrm{Ba_{2}CuO_3}$, and calculate its  electronic band structure.  The obtained band structure is then used to construct multi-orbital models, for which we apply the fluctuation exchange (FLEX)  approximation~\cite{FLEX1,FLEX2} to study the superconductivity.  
We show that $s$-wave and $d$-wave pairings closely compete with 
each other, where we find a peculiar case of coexisting 
intraorbital and interorbital pairings.  
We further show that superconductivity is strongly enhanced if we increase the energy of the $d_{3z^2-r^2}$ band (from its original position obtained by first-principles calculation for $\mathrm{Ba_{2}CuO_3}$) to 
make it ``incipient''~\cite{KurokiArita,inc, inc1, inc2, inc3, inc4, inc9, inc5, inc6, inc7,KobayashiAoki,Matsumoto2018,Misumi,inc8,bi14,Sayyad,Aokireview,bi15,bi16,twoleg_s}, where the lower band edge comes close to the Fermi level within a realistic band filling regime. %70Kを削除
In an even wider scope, 
we reveal that the Lieb model has an intimate relation 
with the two-orbital model of the K$_2$NiF$_4$ structure where a close competition between $s$-wave and $d$-wave pairings is known to occur~\cite{BaRPA}. 
We finally point out a relation between the enhanced superconductivity in the present models and an  enhancement found previously in the bilayer Hubbard model with an incipient band.

\begin{figure}[t]
\centering
\begin{center}
\includegraphics[width=\linewidth]{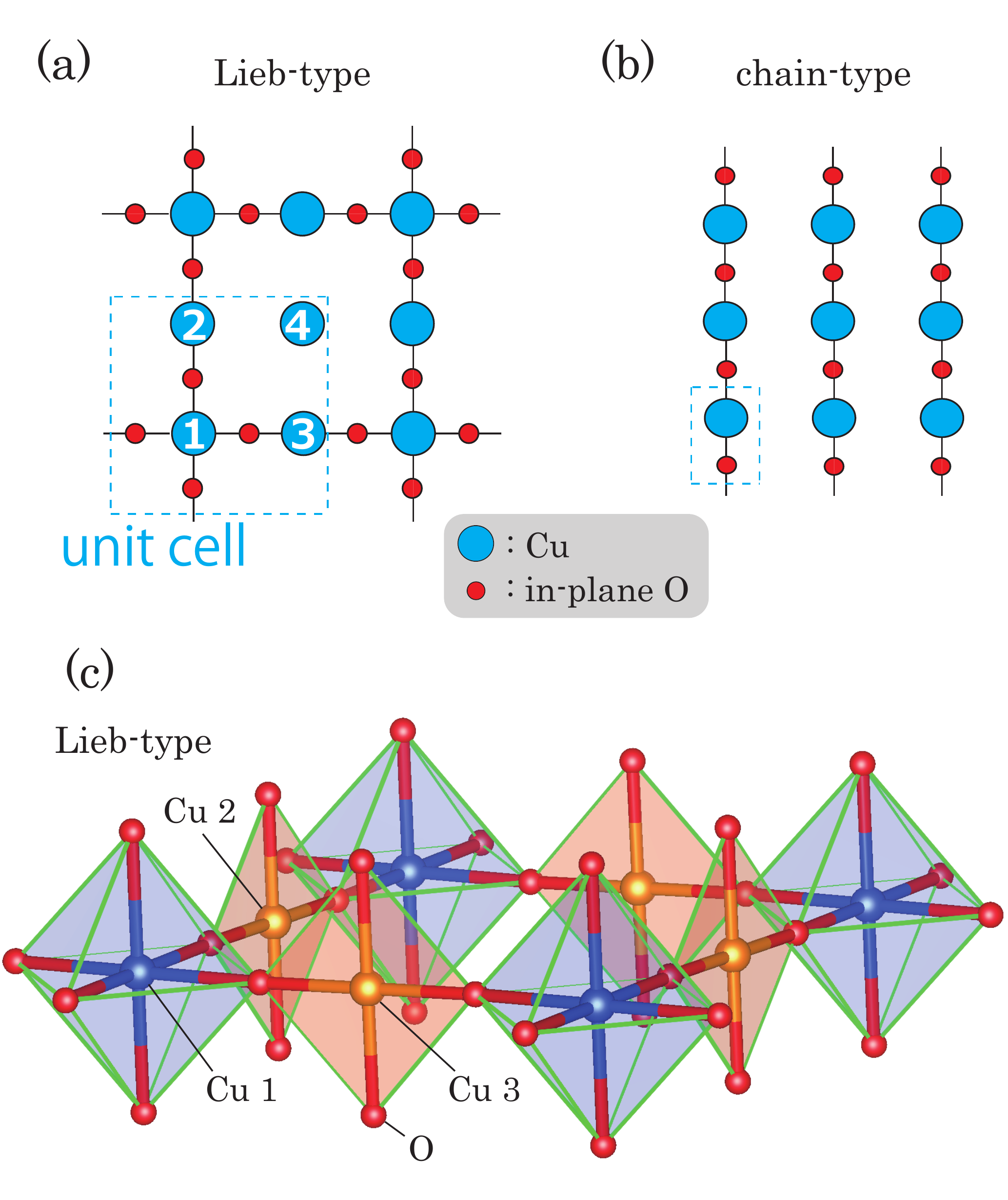}
\caption{Cu-O plane in (a) the Lieb-lattice-type structure, and (b) the chain-type structure. The apical oxygen positions (not displayed) are all occupied in both cases. Ba atoms reside at the same sites as in the K$_2$NiF$_4$-type structure. (c) 
A bird's-eye view of the Lieb structure with VESTA~\cite{VESTA} software. Cu site 4 is not displayed for clarity.}
\end{center}
\label{fig:structure}
\end{figure}

%revision

\section{FORMULATION}
 We consider the Lieb-lattice-type structure for $\mathrm{Ba_{2}CuO_3}$, where the in-plane oxygen deficiencies are ordered as shown in Fig.~1(a).  
Details of the Ba atom position and the unit cell of this structure are 
depicted in Appendix A.  
We perform structural optimization using the Vienna Ab Initio Simulation Package (VASP)~\cite{VASP1,VASP2}. Here, we adopt the generalized gradient approximation formulated by Perdew, Burke, and Ernzerhof for the exchange-correlation energy functional~\cite{PBE} and the projector augmented wave method~\cite{PAW} without the inclusion of the spin-orbit coupling, and take an $8\times8\times8$ $k$-mesh with a plane-wave cutoff energy of $E_{\mathrm{cut}}=650\,\mathrm{eV}$. We also examine the dynamical stability of the Lieb lattice by performing phonon calculation. We employ the finite displacement method as implemented in the PHONOPY software~\cite{phonopy} in combination with VASP. We took a $2\times2\times2$ supercell and a $3\times3\times3$ $k$-mesh. 
%%修正
Other conditions such as the energy cutoff are the same as those adopted in the structural optimization, which is always the case for phonon calculations in general.

For the optimized lattice structure, we obtain the electronic band structure taking a $6\times6\times6$ $k$-mesh  with a plane-wave cutoff energy of $E_{\mathrm{cut}}=550\,\mathrm{eV}$. From the electronic band structure, we extract the maximally localized Wannier functions~\cite{Marzari,Souza} using the WANNIER90 code~\cite{Wannier90}.  
Here we disregard very small hopping parameters to simplify the multiorbital Lieb lattice models (see Appendix B for details).

In order to take account of the electron correlation effects beyond the first principles band calculation, we further introduce the on-site multiorbital interactions with a Hamiltonian,

\begin{equation}
\begin{split}
H_{\mathrm{int}}&=U\sum_{i,\mu}n_{i\mu\uparrow}n_{i\mu\downarrow}\\
&+U'\sum_{i,\mu<\nu,\sigma}n_{i\mu\sigma}n_{i\nu\bar{\sigma}}+(U'-J)\sum_{i,\mu<\nu,\sigma}n_{i\mu\sigma}n_{i\nu\sigma}\\
&-J\sum_{i,\mu\neq\nu}c^{\dagger}_{i\mu\uparrow}c_{i\mu\downarrow}c^{\dagger}_{i\nu\downarrow}c_{i\nu\uparrow}\\
&+J'\sum_{i,\mu\neq\nu}c^{\dagger}_{i\mu\uparrow}c^{\dagger}_{i\mu\downarrow}c_{i\nu\downarrow}c_{i\nu\uparrow}.
\end{split}
\end{equation}
Here,  $i$ denotes the sites, $\mu$, $\nu$ indicate the orbitals, $\sigma$ represents the spins, 
$c^{\dagger}_{i\mu\sigma}$ creates an electron, and $n_{i\mu\sigma}=c^{\dagger}_{i\mu\sigma}c_{i\mu\sigma}$.  
Interactions are $U$, the intraorbital repulsion; $U'$, 
the interorbital repulsion; $J$, Hund's coupling; and $J'$, the pair hopping. We do not consider electron-phonon interactions, since our aim is to investigate an electronic mechanism of superconductivity.
To analyze the many-body effect, here we adopt the FLEX approximation. 

In the FLEX approximation, renormalized Green's function is determined self-consistently 
from the Dyson equation, where the self-energy is calculated by taking 
the bubble and ladder diagrams that consist of the irreducible susceptibility, 
\begin{equation}
\chi^{0}_{l_{1}l_{2}l_{3}l_{4}}(q)=-\frac{T}{N}\sum_{k}G_{l_{3}l_{1}}(k)G_{l_{2}l_{4}}(k+q),
\end{equation}
which is calculated from the renormalized Green's function $G$.
Here $q=(\bm{q},\omega)$ stands for the wave vector $\bm{q}$ 
and the Matsubara frequency $\omega$, $l_{i}$ denotes the orbitals, $T$ is the temperature, and $N$ is the number of $k$ points.
In order to avoid double counting of the effect of the electron interaction already considered in the first principles calculation, we subtract the $\omega=0$ component of the self-energy $\mathrm{Re}\Sigma(\bm{k},0)$ during the self-consistent loop following Ref.~\cite{difsigma}.
Hence, the Fermi surface of the models remains unchanged even after the correlation effects are taken into account by the FLEX calculation.
%追加%追加
It should be noted that the double counting is not rigorously avoided since $\Sigma(k,\omega=0)$ in FLEX is not the same as that in the DFT calculation.

Using the obtained Green's function along with 
the spin ($\hat{\chi}_{\rm s}$) and charge ($\hat{\chi}_{\rm c}$) susceptibilities,  
\begin{equation}
\hat{\chi}_{\rm s}(q)=\frac{\hat{\chi}^{0}(q)}{1-\hat{S}\hat{\chi}^{0}(q)},
\end{equation}
\begin{equation}
\hat{\chi}_{\rm c}(q)=\frac{\hat{\chi}^{0}(q)}{1+\hat{C}\hat{\chi}^{0}(q)},
\end{equation}
which are matrices for multiorbital systems 
with the interaction matrices given as
\begin{equation}
S_{l_{1}l_{2}l_{3}l_{4}}=\begin{cases}
U, & l_{1}=l_{2}=l_{3}=l_{4},\\
U', & l_{1}=l_{3}\neq l_{2}=l_{4},\\
J, & l_{1}=l_{2}\neq l_{3}=l_{4},\\
J', & l_{1}=l_{4}\neq l_{2}=l_{3},
\end{cases}
\end{equation}
\begin{equation}
C_{l_{1}l_{2}l_{3}l_{4}}=\begin{cases}
U, & l_{1}=l_{2}=l_{3}=l_{4},\\
-2U'+J, & l_{1}=l_{3}\neq l_{2}=l_{4},\\ %修正
2U'-J, & l_{1}=l_{2}\neq l_{3}=l_{4},\\
J', & l_{1}=l_{4}\neq l_{2}=l_{3},
\end{cases}
\end{equation}
we obtain the effective spin-singlet pairing interaction, 
\begin{equation}
\hat{\Gamma}(q)=\frac{3}{2}\hat{S}\hat{\chi}_{\rm s}(q)\hat{S}-\frac{1}{2}\hat{C}\hat{\chi}_{\rm c}(q)\hat{C}+\frac{1}{2}(\hat{S}+\hat{C}),
\end{equation}
which is plugged into the linearized Eliashberg equation,
\begin{equation}
\begin{split}
\lambda\Delta_{\mu\nu}(k)=&-\frac{T}{N}\sum_{q,m_{i}}\Gamma_{\mu m_{1}m_{4}\nu}(q)G_{m_{1}m_{2}}(k-q)\\
&\times\Delta_{m_{2}m_{3}}(k-q)G_{m_{4}m_{3}}(q-k),
\end{split}
\end{equation}
where the gap function $\Delta_{\mu\nu}(k)$ is also a matrix.  
%追加
$G$ in Eq.(8) is the renormalized Green's function obtained from the FLEX calculation. Through $G$, the mass renormalization and finite-lifetime effects are taken into account.
The maximum eigenvalue $\lambda$ of this equation reaches unity at $T=T_{c}$, 
so that $\lambda$ calculated at a fixed temperature can be a measure of $T_c$. Throughout the present study, we calculate $\lambda$ at $T=0.01\,\mathrm{eV}$. 
We refer to the eigenfunction of the linearized Eliashberg equation as the gap function.
%追加
Note that since Eq.(8) is a linearized equation, the absolute value of the gap function does not have any physical meaning, and only its relative magnitude among different orbital components and its symmetry are relevant.
Both the Green's functions and the gap functions are obtained first in the orbital representation, which can be transformed into the band representation with a unitary transformation.  
Green's functions will be presented by taking its absolute value. Also, 
Green's functions  and the gap functions will be presented for the lowest  Fermionic Matsubara frequency $i\pi k_{\mathrm{B}}T$, and the effective pairing interactions $\Gamma_{\alpha\beta\beta'\alpha'}$ will be presented at the lowest bosonic Matsubara frequency 0.  

Assuming a rigid band obtained for the Lieb-lattice-type model, we vary the band filling in a regime that contains a case corresponding to $\mathrm{Ba_{2}CuO_{3+\delta}}$ with a realistic $\delta\sim 0.2$. 
In the calculation, we take $2048$ Matsubara frequencies and a $16\times16\times2$ $k$-point mesh.
%追加
We have checked that calculation taking $32\times32\times2$ $k$-point mesh gives essentially the same results.

\section{\label{sec:level3}RESULTS}
\subsection{Stability of the Lieb structure for Ba$_{2}$CuO$_{3}$}
We start with the stability of the Lieb-lattice-type structure for $\mathrm{Ba_{2}CuO_3}$. 
The obtained total energy of the optimized lattice structure is $E_{\mathrm{tot}}(\mathrm{Lieb})=-33.95\,\mathrm{eV}$. For comparison, we have also performed structural optimization for the chain-type structure of $\mathrm{Ba_{2}CuO_3}$, shown in Fig.~1(b) (see Appendix A for the actual structure), whose total energy is estimated as $E_{\mathrm{tot}}(\mathrm{chain})=-33.99\,\mathrm{eV}$. 
%修正修正%%
Thus, the total energies of the two structures turn out to be quite close to each other; considering the accuracy of the first principles calculation, a difference of 40\,\,meV can be reversed, e.g.,~by the effects of the correlation and/or excess oxygens not taken into account here~\cite{MnO}. Then, given the fact that the chain-type structure is known to be realized in an existing material Sr$_{2}$CuO$_{3}$~\cite{Srchain1,Srchain2,Srchain3} but inconsistent with the tetragonal symmetry, the Lieb-lattice-type structure may be considered as a realistic candidate for the lattice structure of Ba$_{2}$CuO$_{3+\delta}$.
We further calculate the phonon dispersion as presented in Fig.~2. 
We find that no imaginary phonon modes are present for this lattice structure, which suggests dynamical stability of the present Lieb-lattice structure.

As for the apical oxygen position determined by structural optimization, its distance measured from the in-plane Cu site turns out to 
depend on the site: 2.18\,{\AA} above Cu site 1, 1.96\,{\AA} above Cu sites  2 and 3, and 1.82\,{\AA} above Cu site 4. Namely, the sites with smaller oxygen coordination numbers have lower apical oxygen heights. The average value is 1.98\,{\AA}, which is substantially smaller than the value (2.42\,{\AA}) for La$_2$CuO$_4$.  
We may note that this is qualitatively consistent with the experimental value 
of 1.86\,{\AA}~\cite{Ba}, if we consider the fact that the excess oxygens in the actual material will increase in-plane holes that should attract the apical oxygens.
%追加
In fact, for a cuprate La$_{2-x}$Sr$_{x}$CuO$_{4}$, it is actually observed experimentally in Ref.~\cite{apicalhole} that hole doping by substituting La with Sr (which has nearly the same ion radius as La) induces reduction of the apical oxygen height from 2.42\,{\AA} ($x=0$) to 2.30\,{\AA} ($x=0.2$).

\begin{figure}[t]
\centering
\begin{center}
\includegraphics[width=\linewidth]{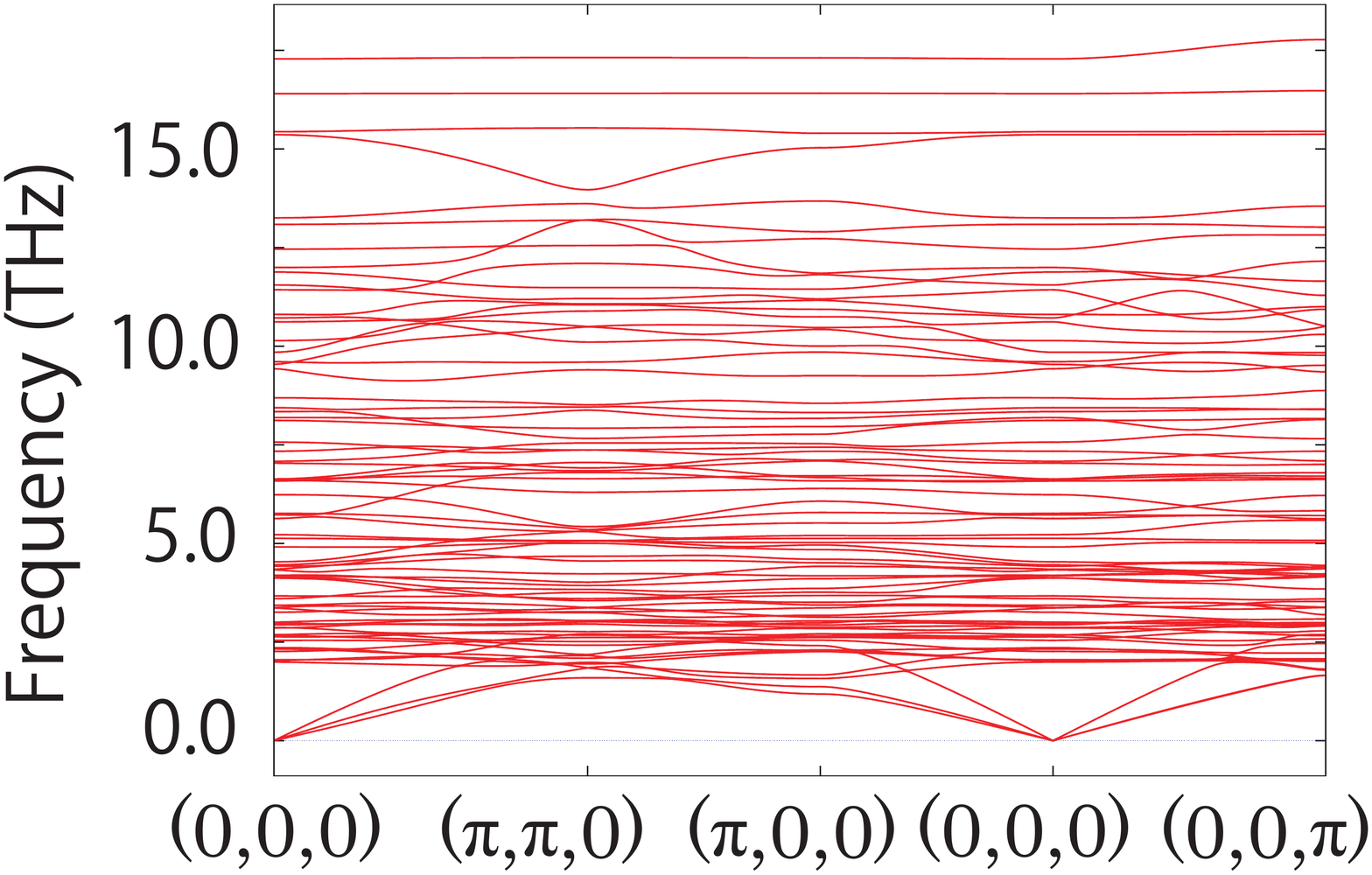}
\caption{Phonon dispersion for the Lieb-lattice-type $\mathrm{Ba_{2}CuO_{3}}$.}
\end{center}
\label{fig:Lieb_phonon}
\end{figure}

\subsection{Electronic band structure of the Lieb lattice}
From the optimized crystal structure, we obtain the electronic band structure as presented in Fig.~3, where we also display the weight of the Cu $d_{x^2-y^2}$ and Cu $d_{3z^2-r^2}$ orbital characters. 
The bands originating from the orbitals at Cu site 4, which has no neighboring oxygens in 
the plane, have energies very low, so that site 4 is irrelevant 
in the model building. We then extract tight-binding models 
downfolded in terms of the maximally localized Wannier orbitals. We consider 
the two $e_g$ orbitals centered at each of the Cu sites 1, 2, and 3, 
which results in a six-orbital model, where the oxygen orbitals are implicitly taken into account through the Wannier orbitals.  
The tight-binding  band structure is shown in Fig.~4(a) (blue lines), 
which accurately coincides with the first principles bands.  
The Wannier orbitals are depicted in Fig.~4(c).  
In the six-orbital band structure, we notice that the two lowermost bands are 
nearly flat, which originates from the orbitals of sites 2 and 3 
pointing to the direction where oxygens are absent [orbitals 3 and 5 in Fig.~4(c)]. The two bands in the middle are the bonding bands originating from site 1 $d_{x^2-y^2}$ [Fig.~4(c), orbital 1] and $d_{3z^2-r^2}$ [Fig.~4(c), orbital 2], hybridized with the orbitals of sites 2 and 3 extended toward site 1 and also toward the apical oxygens [Fig.~4(c), orbitals 4 and 6]. An important point here is that the low apical oxygen height makes the upper (lower) bonding band mainly originated from the $d_{3z^2-r^2}$ ($d_{x^2-y^2}$) at site 1, because the energy levels of these orbitals are {\it inverted} from those in the conventional cuprates, as suggested in previous studies~\cite{Ba,BaRPA,LeJHu}.  
 The top two bands are the antibonding bands from the hybridization between site 1 and sites 2 and 3. Here, the orbitals at sites 2 and 3, hybridized with site 1 orbitals, have energy higher than site 1 $d_{x^2-y^2}$ and $d_{3z^2-r^2}$ orbitals because the apical oxygens at sites 2 and 3 are closer to Cu.

In the actual material $\mathrm{Ba}_{2}\mathrm{CuO}_{3+\delta}$, 
the Fermi level should 
be shifted downward to intersect the middle two bands because the oxygen content is larger than in $\mathrm{Ba_{2}CuO_3}$. Therefore, we further construct a two-orbital model that extracts the two middle bands, as shown in Fig.~4(b), by considering $d_{x^2-y^2}$ and $d_{3z^2-r^2}$ Wannier functions centered at Cu site 1. In this model, the orbitals of Cu sites 2 and 3, as well as the oxygen $2p$ orbitals, are implicitly taken into account through the  Wannier functions. In this two-orbital model, we shall sometimes refer to the $d_{x^2-y^2}$ and $d_{3z^2-r^2}$ Wannier orbitals as orbitals 1 and 2, respectively.
%追加
Note that, although we cannot rule out the possibility of a high-spin state, in the following analysis we assume that the ground state is in a low-spin state.

We vary the band filling in the following FLEX calculation assuming a rigid band in these models, where the correspondence between the oxygen content and the band filling is as follows. In $\mathrm{Ba_{2}CuO_{3}}$, the nominal Cu valence is $+2$, so the electron configuration is $3d^{9}$ on average, that is, (three $e_g$ electrons) $\times$ (four sites) $=$ (12 $e_g$ electrons) per unit cell. Since the $e_g$ orbitals at Cu site 4 are fully occupied by electrons, there are $8$ $e_g$ electrons per unit cell in the six-orbital model. Namely, the band filling $n$, defined as the number of electrons per spin per unit cell, is $n=4$. Similarly, $n=3$ corresponds to $\mathrm{Ba_{2}CuO_{3.25}}$. The band filling of the six-orbital model 
subtracted by two gives that of the two-orbital model because the bottom two 
occupied bands are ignored in the latter; namely, $n=1$ and $n=2$ in the two-orbital model correspond to $\mathrm{Ba_{2}CuO_{3.25}}$ and  $\mathrm{Ba_{2}CuO_{3}}$, respectively.  
An interesting point here is that the oxygen content of O$_{3.25}$, 
which is close to the actual experimental situation of $\sim$ O$_{3.2}$~\cite{Ba} and implies a large amount of hole doping ($\sim 50$\%) in the usual sense of the term, 
in fact corresponds to half filling
in the present two-orbital model of the Lieb-type lattice. We shall indeed see
that the electron correlation effect are maximized and thus superconductivity is optimized around this band filling.

For later reference, we have also obtained, via structural optimization, the electronic band structure for the K$_2$NiF$_4$-structured Ba$_2$CuO$_4$  by VASP taking a $8\times8\times8$ $k$-mesh with a plane-wave cutoff energy of $E_{\mathrm{cut}}=550\,\mathrm{eV}$ [Figs.~3(d) and (e)]. There, we construct a two-orbital model by extracting the $e_g$ orbitals centered at the Cu sites [Fig.~4(d)].  We shall discuss the relation between the Lieb lattice-type and K$_2$NiF$_4$-type structures in the Discussions section.

%DOS追加
\begin{figure}[t]
\centering
\begin{center}
\includegraphics[width=\linewidth]{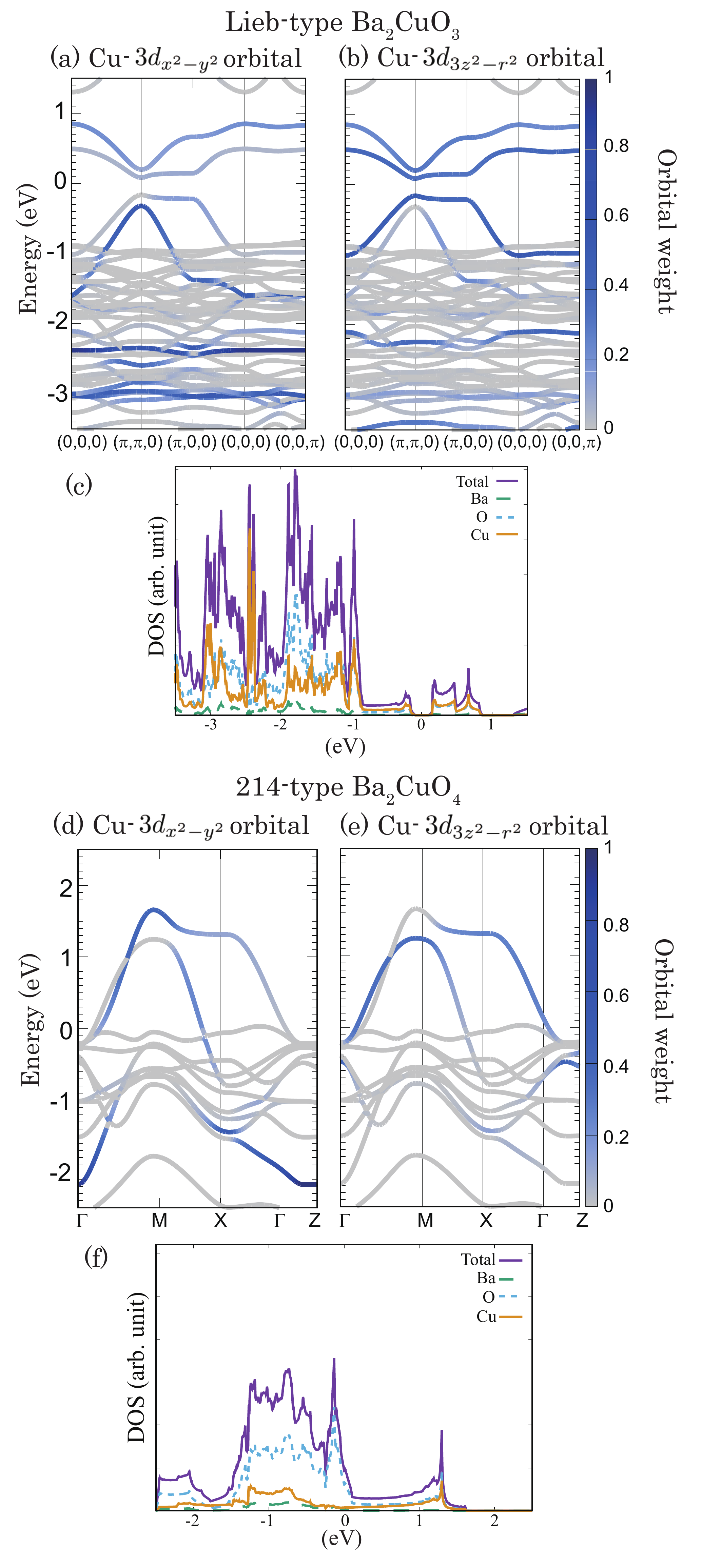}
\caption{Upper panels: first principles band structure [(a), (b)] and (projected) DOS (c) obtained for the Lieb-lattice-type structure for $\mathrm{Ba_{2}CuO_3}$.  Lower panels: first principles band structure [(d), ,(e)] and (projected) DOS (f) of $\mathrm{Ba_{2}CuO_4}$ with the $\mathrm{K}_{2}\mathrm{NiF}_{4}$-type structure. Blue lines represent the weight of  the Cu $d_{x^{2}-y^{2}}$ character [(a), (c)] or  Cu $d_{3z^{2}-r^{2}}$ character [(b), (d)].}
\end{center}
\label{fig:band1}
\end{figure}

\begin{figure*}[t]
\centering
\begin{center}
\includegraphics[width=\linewidth]{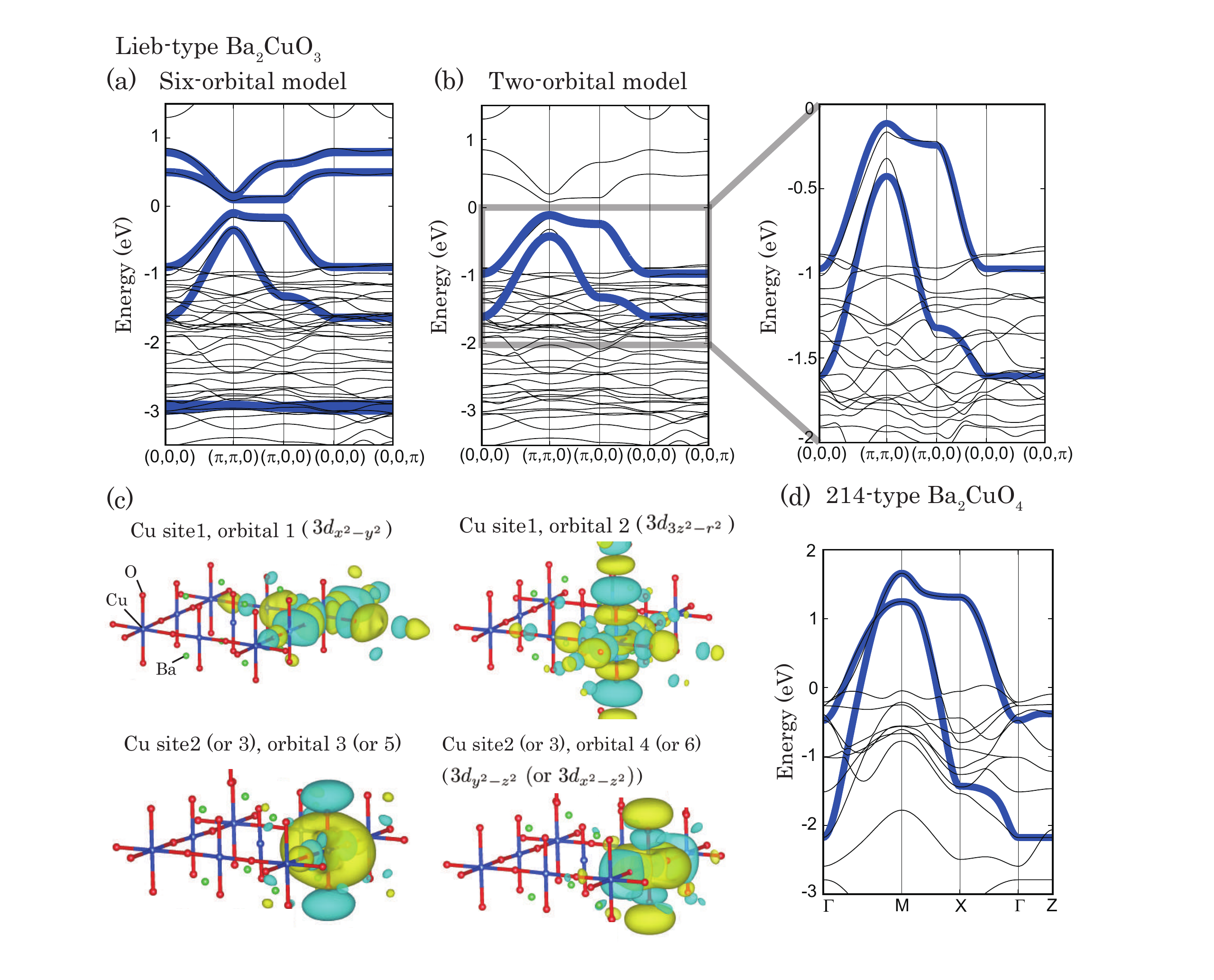}
\caption{Band structure (blue lines) of the (a) six-orbital model and (b) two-orbital model of the Lieb lattice, with an entanglement on the right.  
Superposed is the first principles band structure (black lines). 
(c) Wannier orbitals in the six-orbital model with VESTA software. (d) Band structure of the two-orbital model of the $\mathrm{K}_{2}\mathrm{NiF}_{4}$-type $\mathrm{Ba_{2}CuO_4}$ superposed to the first-principles band structure.}
\end{center}
\label{fig:wannier}
\end{figure*}

\subsection{Superconductivity}
We now move on to the FLEX calculation for superconductivity. 
We start with the two-orbital model for the Lieb lattice. 
Figure 5 plots the eigenvalue of the Eliashberg equation $\lambda$ at $T=0.01$\,eV for the $s$- and $d$-wave pairing symmetries.  
We can see that the two pairing symmetries give somewhat close values of $\lambda$, where the $s$ wave 
slightly dominates within the parameter regime studied. In both symmetries, $\lambda$ is maximized at $n=1$. 
This is because electron correlation effects are maximized around half-filling, and, as stressed above, we can notice this band filling 
corresponds to the oxygen content of O$_{3.25}$, close to the actual material.

Before we go any further, we have to carefully examine the gap functions and the definition of the pairing symmetry.  
Figure 6 displays the gap functions in both orbital and band representations.  
Note above all that the gap function in the present multi-orbital system 
is a matrix.  A curious finding in 
Fig.~6 is that the pairing symmetry is 
inverted between diagonal and off-diagonal 
matrix elements, i.e., $s$-wave diagonal elements 
are accompanied by $d$-wave off-diagonal ones [Figs.~6(a) and (c)], 
while $d$-wave diagonal elements 
are accompanied by $s$-wave off-diagonal ones [Figs.~6(b) and (d)].  
This occurs both in orbital  [Figs.~6(a) and (b)] and band  [Figs.~6(c) and (d)] representations. 
We can trace the curious phenomenon back to 
the hybridization between the $d_{x^2-y^2}$ and $d_{3z^2-r^2}$ orbitals, where the hybridization has sign structure in a $d_{x^2-y^2}$ symmetry.  
In Fig.~5, we have abbreviated the diagonal $s$-wave 
with off-diagonal $d$-wave as $s$-wave, 
and the diagonal $d$ 
with off-diagonal $s$ as $d$-wave. We will adopt this abbreviation hereafter.

If we focus on the orbital-diagonal elements of the gap function, the $d$-wave has the $\cos(k_x)-\cos(k_y)$ form usually encountered in similar analysis of the conventional cuprates. In real space, this corresponds to a nearest-neighbor pairing, whose wave function changes its sign upon 90-degree rotation. On the other hand, the $s$-wave gap roughly has the form $\cos(k_x)+\cos(k_y)$, which implies that this is 
basically an extended $s$-wave with a pair 
residing on nearest-neighbor sites. We can compare the band representation of the $s$-wave gap function with Green's function $|G(\bm{k})|$ in Figs.~6(e) and (f).   The ridges in $|G(\bm{k})|$ represent the Fermi surface, and 
we have two pieces for the Fermi surface in this 
two-orbital model as seen from $|G_{11}|$ and $|G_{22}|$.  We can then realize that 
the gap has a sign-inversion across the two Fermi surfaces; namely, we have here the so-called $s\pm$-wave gap function, as depicted in the left panel of Fig.~7(b), which is reminiscent of the iron-based superconductors~\cite{inc1,iron_review} as far as the Fermi surfaces are concerned.

If we now turn to the orbital-off-diagonal elements of the gap function, the symmetries ($s$ or $d$) are exchanged from 
the diagonal elements as we have noted, but we can also 
notice in Fig.~6 that the amplitudes of the off-diagonal 
elements are comparable to those of the diagonal ones, 
which indicates that the inter-orbital pairing has significant contributions.  Now, inter-orbital spin-{\it triplet} pairing has been studied in the multiorbital Hubbard model 
that has degenerate orbitals~\cite{inter,inter2,inter3,inter4}, but in the present case the gap function has the  symmetry $\Delta_{\alpha\beta}(\bm{k})=\Delta_{\beta\alpha}(-\bm{k})$ that signifies a {\it singlet} pairing.  
We can intuitively grasp, in real space, 
the coexistence of the inter- and intra-orbital pairings in Fig.~7(a), where both of intra-orbital and inter-orbital pairs reside on nearest neighbors, as we have explained.
Since pairing between electrons having large energy difference is unlikely, this kind of pairing is peculiar to systems where the two orbitals are close in energy.

We now grasp how the coexisting inter- and intra-orbital pairings arise. We find, in the typical parameter regime considered, that the diagonal components of Green's function (at the lowest Matsubara frequency) are larger than the off-diagonal ones, with their real part much larger than the imaginary part, and that 
they satisfy $\mathrm{Re}[G_{mm}(k)]\mathrm{Re}[G_{ll}(-k)]>0$, where $m,l=$orbital 1 $(d_{x^2-y^2})$ or 2 $(d_{3z^2-r^2})$. We can thus roughly extract the contributions 
of these components in the linearized Eliashberg equation as
\footnotesize
\begin{eqnarray}
\lambda\Delta_{11}(k)\sim-\Gamma_{1111}(q)G_{11}(k-q)\Delta_{11}(k-q)G_{11}(q-k),\\
\label{eq:Gamma1111}
\lambda\Delta_{12}(k)\sim-\Gamma_{1212}(q)G_{22}(k-q)\Delta_{21}(k-q)G_{11}(q-k),\\
\label{eq:Gamma1212}
\lambda\Delta_{11}(k)\sim-\Gamma_{1221}(q)G_{22}(k-q)\Delta_{22}(k-q)G_{22}(q-k),\\
\label{eq:Gamma1221}
\lambda\Delta_{12}(k)\sim-\Gamma_{1122}(q)G_{11}(k-q)\Delta_{12}(k-q)G_{22}(q-k),
\label{eq:Gamma1122}
\end{eqnarray}
\normalsize

for the intraorbital [Eqs.(9) and (12)] and interorbital [Eqs.(10) and (11)] pair scattering channels with the Feynman diagrams for the pairing interaction $\Gamma_{\alpha\beta\beta'\alpha'}$ as depicted in Figs.~8(i) and 8(j) and Figs.~9(i) and 9(j).  See Table I. \\

\begin{table}
\caption{The pairing interactions.}
\begin{tabular}{c|c|c} 
&  pairing & pair scattering \\ \hline
$\Gamma_{1111}, \Gamma_{2222}$ & intra-orbital & intra-orbital \\ 
$\Gamma_{1122}$ & inter-orbital & intra-orbital \\
$\Gamma_{1221}$ & intra-orbital & inter-orbital \\
$\Gamma_{1212}$ & inter-orbital & inter-orbital 
\end{tabular} \\
\end{table}
%%%%

To identify which interaction parameters in the 
Hamiltonian dominate these pairing interactions, we calculate the eigenvalue $\lambda$ against the band filling for various choices of $(U,U',J,J')$ as shown in Fig.~10. Here we permit breaking the orbital rotational 
symmetry ($U'=U-2J$) in order to extract the effect of each interaction. The pairing interaction $\Gamma_{\alpha\beta\beta'\alpha'}$ for the varied interactions 
is depicted in Figs.~8 and 9.  
We find in Fig.~10 that increasing the inter-orbital interactions, $U'$, $J$, and $J'$, enhances $\lambda$.  On the other hand, Fig.~10(c) shows increasing the intra-orbital 
$U$ initially enhances $\lambda$, which is rounded 
off for larger $U$. If we compare this with Figs.~9(a)--(c) and 9(e)--(g), we reveal that increasing $U'$, $J$, and $J'$ enhances $\Gamma_{1221}$  and $\Gamma_{1212}$, which in turn enhances $\lambda$. On the other hand, increasing $U$ enhances $\Gamma_{1111}$ [Figs.~8 (a) and (d)] and $\Gamma_{1122}$ [Figs.~8(e) and (h)], but suppresses $\Gamma_{1221}$ [Figs.~9(a) and (d)] and $\Gamma_{1212}$ [Figs.~9(e) and (h)].  
The increase of $\Gamma_{1111}$ 
enhances the intra-orbital pairing while the 
increase of $\Gamma_{1122}$ enhances the interorbital pairings, but the suppression of $\Gamma_{1221}$ degrades intra-orbital pairings 
while the suppression of $\Gamma_{1212}$ degrades inter-orbital pairings, which is probably the origin of the nonmonotonic behavior against the $U$ variation. 

A salient feature here is that $\Gamma_{\alpha\beta\beta'\alpha'}$s all have peaks  around $\Vec{q} = (\pi,\pi)$. 
From Eqs.~(9)--(\ref{eq:Gamma1122}), the portions of the gap function in the regions where Green's function is large 
should change sign across $(\pi,\pi)$ within $\Delta_{11}$, $\Delta_{12}$, and between $\Delta_{11}$ and $\Delta_{22}$ as indicated by yellow arrows in Fig.~6(a). Which of the $s$- or $d$-wave pairings dominates should depend on the shape of the Fermi surface; if we reduce the inter-orbital hopping, the Fermi surface of the lower band is less warped as depicted in Fig.~7(b), and the Fermi surface approaches $(\pi,0)/(0,\pi)$, which favors $d$-wave pairing because $\cos(k_x)-\cos(k_y)$ has large amplitudes around $(\pi,0)/(0,\pi)$. In the case without the interorbital hopping ($t_{12}=0$), however, we find that the value of $d$-wave $\lambda$ ($\sim0.1$, not shown)  is smaller than that of $s$-wave $\lambda$ for the original value of $t_{12}$. Also, in the absence of $t_{12}$, the off-diagonal component,  $\Delta_{12}$, in the orbital representation of the gap function vanishes. Therefore, we arrive at a mechanism in which 
the effect of the interorbital hybridization can enhance the superconductivity through the {\it coexisting intra- and interorbital pairings}.

\begin{figure}[b]
\centering
\begin{center}
\includegraphics[width=\linewidth]{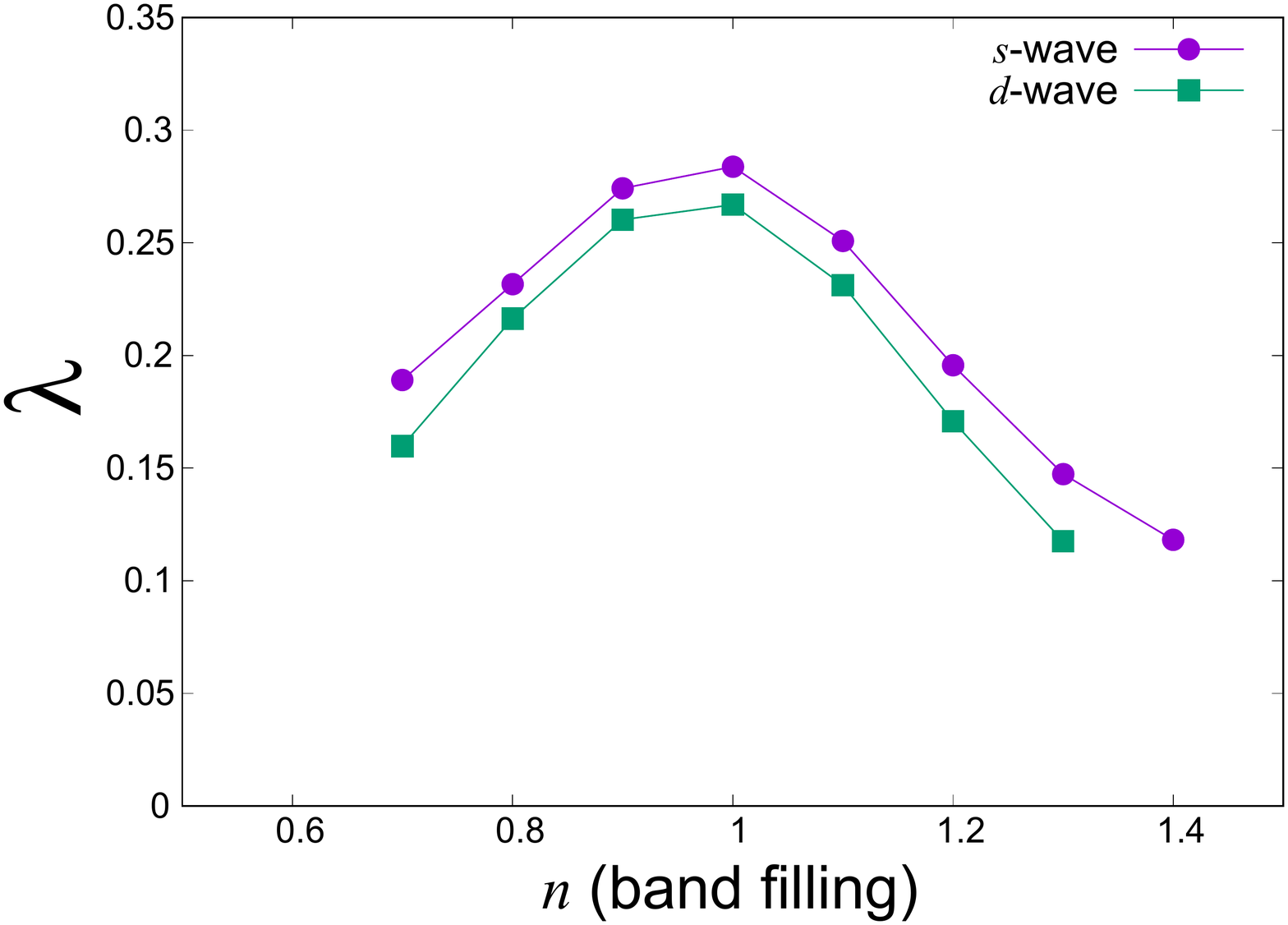}
\caption{Eigenvalue $\lambda$ of the two-orbital model for $s$-wave (intraorbital $s$ and interorbital $d$) and $d$-wave (intra-orbital $d$ and inter-orbital $s$) pairings plotted against the band filling. The parameter values adopted are $U=2.0\,\mathrm{eV}$, $J=J'=U/10$, and $U'=U-2J$. 
%追加
In the two-orbital model, we use $U=2$~eV, which is somewhat smaller than that of the conventional cuprates~\cite{Sa40,Sa41,Sa42}, to take into account the wide spread of the Wannier functions across neighboring Cu atoms.
}
\end{center}
\label{fig:svsd}
\end{figure}

\begin{figure*}[t!]
%\centering
\begin{center}
\includegraphics[width=0.9\linewidth]{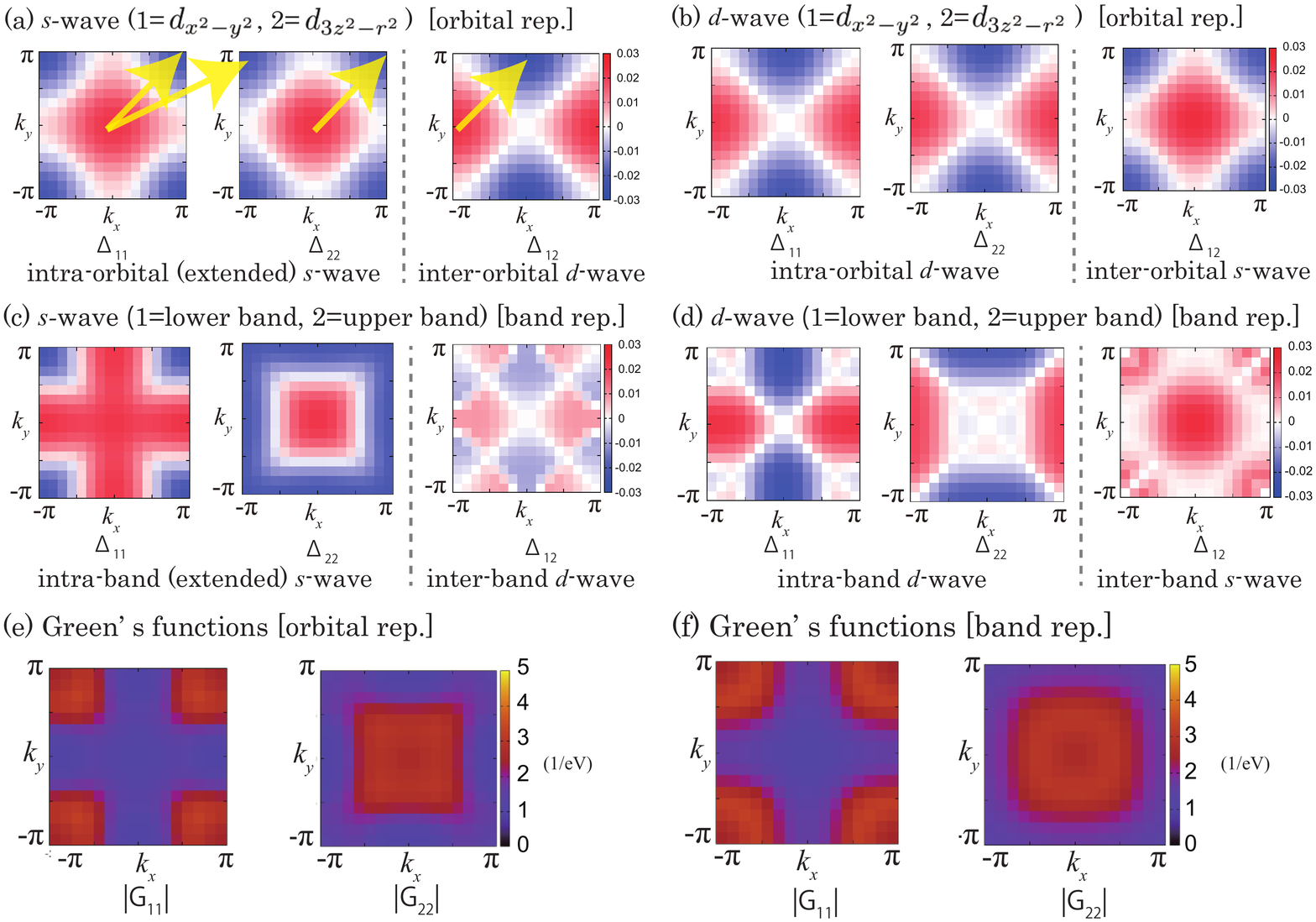}
\caption{For the two-orbital model, 
(a) the intra-orbital $s$-wave with an inter-orbital $d$-wave 
gap functions.  
The portions of the gap functions  that change sign across the wave vector $(\pi,\pi)$ are indicated by yellow arrows (see the text). 
(b) The intra-orbital $d$-wave with an 
inter-orbital $s$-wave gap functions are displayed 
in  the orbital representation.  
Panels (c) and (d) represent them in the band representation, respectively.  
%追加
Note that since Eq.(8) is a linearized equation, the absolute value of the gap function does not have any physical meaning, and only its relative magnitude among different orbital components and its symmetry are relevant.
(e) The absolute value of Green's function in the orbital representation, 
while (f) shows them in the band representation.  
The parameter values adopted are $n=1.0$, $U=2.0\,\mathrm{eV}$, $J=J'=U/10$, and $U'=U-2J$.}
\end{center}
\label{fig:sdgap}
\end{figure*}

\begin{figure}[h]
\centering
\begin{center}
\includegraphics[width=\linewidth]{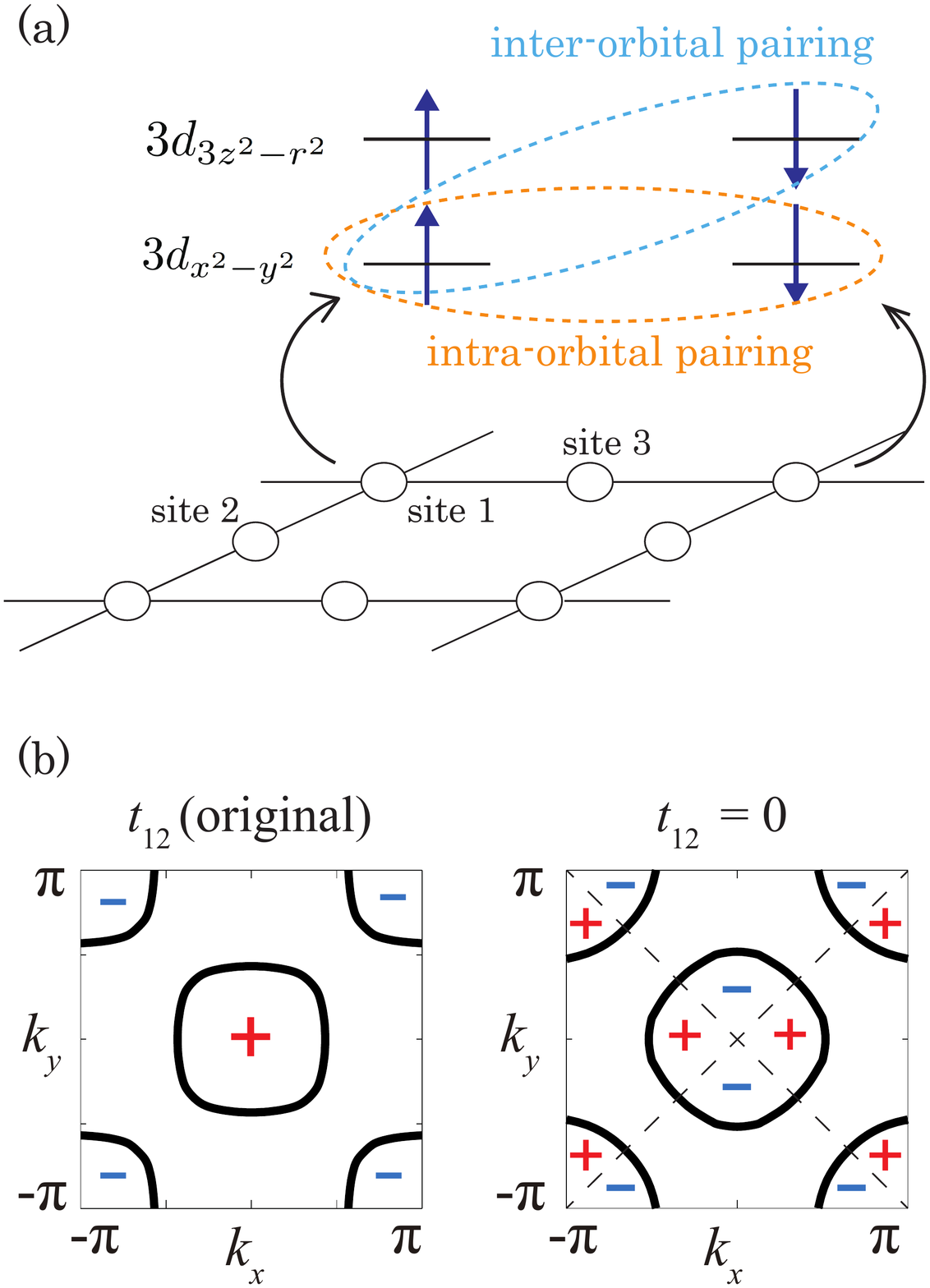}
\caption{(a) Schematics  of the coexisting intra- and inter-orbital nearest-neighbor pairings. In the upper figure, we depict the energy level of the two orbitals centered at site 1 of the Lieb-type lattice, and the electrons occupying those orbitals and forming nearest-neighbor pairs. Note that  ``nearest neighbor”  here refers to the 
nearest-neighbor unit cells rather than the sites.
(b) The Fermi surface at $n=1.0$ for the original band structure of the two-orbital model (left) and that of the band structure in the absence of the inter-orbital hopping (right). The signs in the gap functions 
are indicated for the $s\pm$-wave (left) and $d$-wave (right).}
\end{center}
\label{fig:pair}
\end{figure}

\begin{figure}[h!]
\centering
\begin{center}
\includegraphics[width=0.935\linewidth]{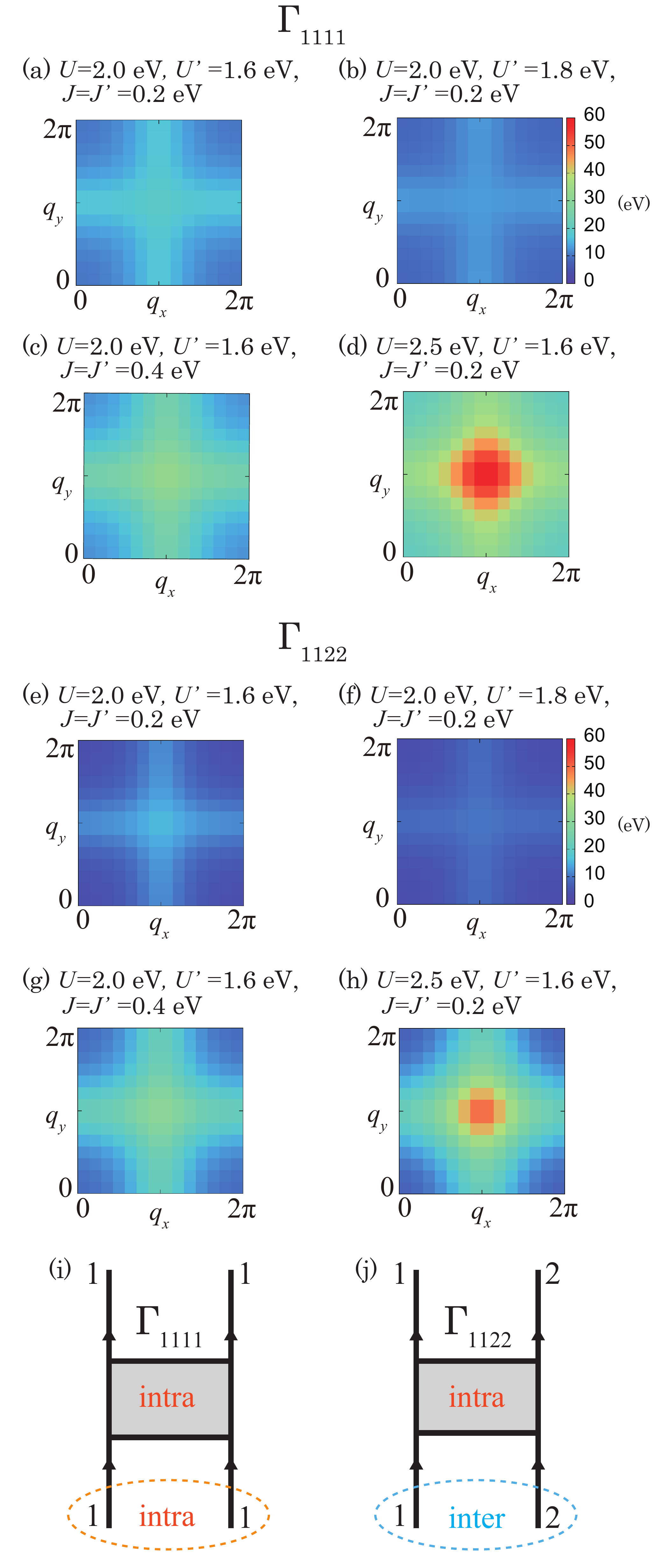}
\caption{The interaction dependence of the effective 
intra-orbital interactions, $\Gamma_{1111}$ [(a)--(d)], 
and inter-orbital interactions, $\Gamma_{1122}$ [(e)--(h)], of 
intra-orbital pairs at the lowest Matsubara frequency.  
The bottom panels depict the Feynman diagram of  $\Gamma_{1111}$ (i) 
and $\Gamma_{1122}$ (j). Note that the pairing interactions are plotted 
over the range $0\leq q_x, q_y \leq 2\pi$ to display the peak structure around $(\pi,\pi)$ clearly.}
\end{center}
\label{fig:G1111}
\end{figure}

\begin{figure}[h!]
\centering
\begin{center}
\includegraphics[width=0.935\linewidth]{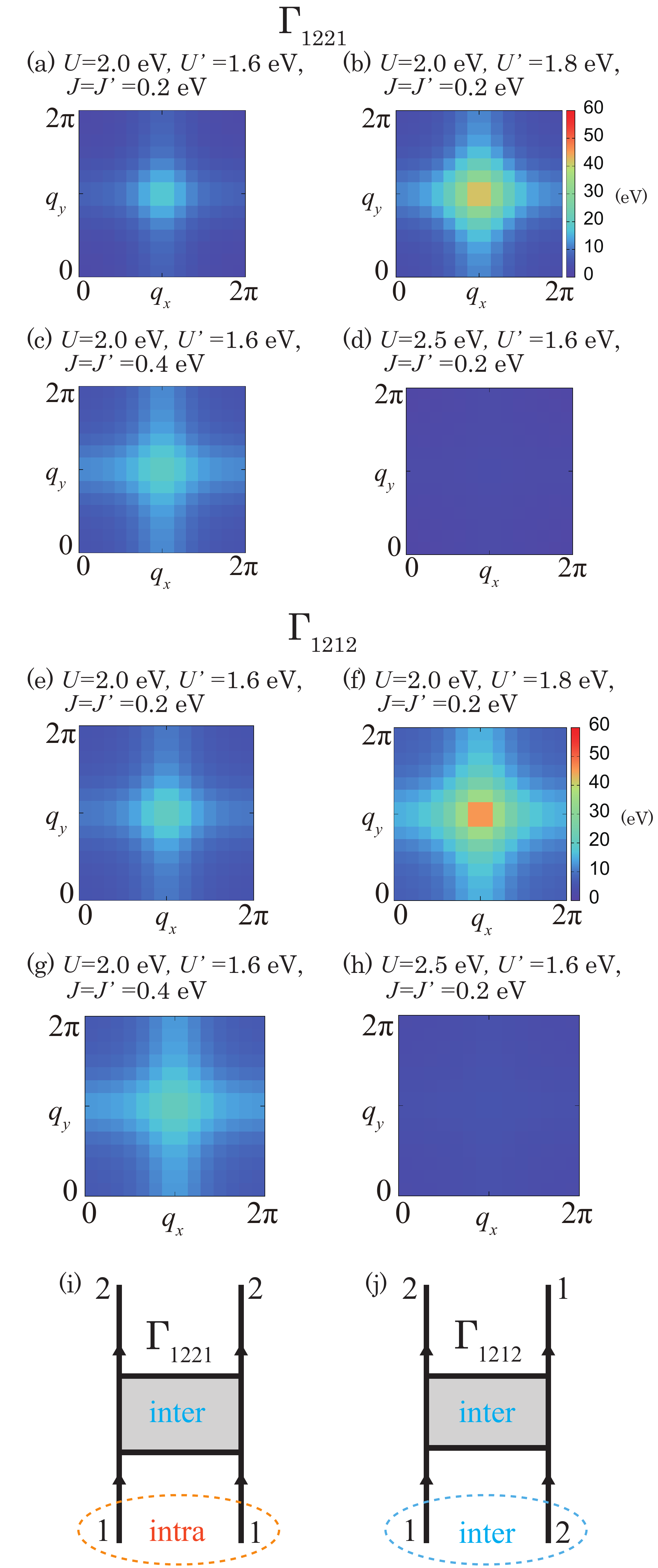}
\caption{The interaction dependence of the effective 
inter-orbital interactions  of 
intra-orbital pairs, $\Gamma_{1221}$ [(a)--(d)], 
and of inter-orbital pairs, $\Gamma_{1212}$ [(e)--(h)], at the lowest Matsubara frequency.  The bottom panels depict the Feynman diagram of $\Gamma_{1221}$ (i) and $\Gamma_{1212}$ (j).  
Note that the pairing interactions are plotted 
over the range $0\leq q_x, q_y \leq 2\pi$ to display the peak structure around $(\pi,\pi)$ clearly.}
\end{center}
\label{fig:G1212}
\end{figure}

\begin{figure}[b]
\centering
\begin{center}
\includegraphics[width=\linewidth]{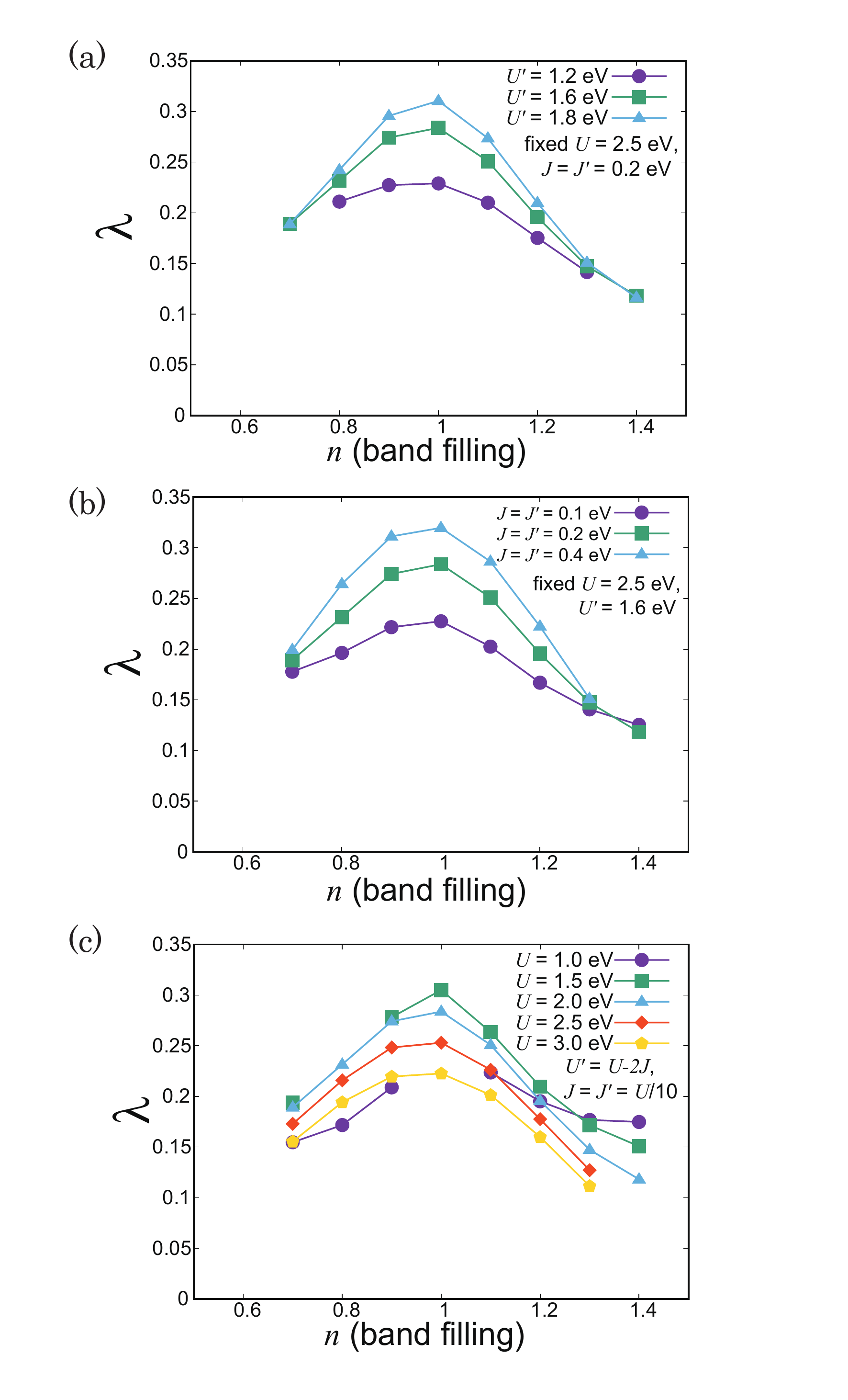}
\caption{$s$-wave eigenvalue $\lambda$ plotted against the band filling  in the two-orbital model for the interaction values varied in different ways. (a) $U'$ dependence with fixed $U=2.0\,\mathrm{eV}$ and $J=J'=0.2\,\mathrm{eV}$, (b) $J$, $J'$ dependence with fixed $U=2.0\,\mathrm{eV}$ and $U'=1.6\,\mathrm{eV}$, 
and (c) $U$ dependence with $U'=U-2J$ and $J=J'=U/10$.}
\end{center}
\label{fig:interaction_dep}
\end{figure}

Finally, let us turn to the six-orbital model in its FLEX results. We show the band filling dependence of $\lambda$ in Fig.~11, and  the gap functions and Green's functions in Fig.~12, for $s$- and $d$-wave pairings. We find that the results are similar to those obtained for the two-orbital model. Namely, $s$-wave and $d$-wave  closely compete with each other, and the intra- and inter-orbital pairing components coexist. Further understanding of the relation between the two-orbital and six-orbital models is given in the Discussions section below.

\begin{figure}[t]
\centering
\begin{center}
\includegraphics[width=\linewidth]{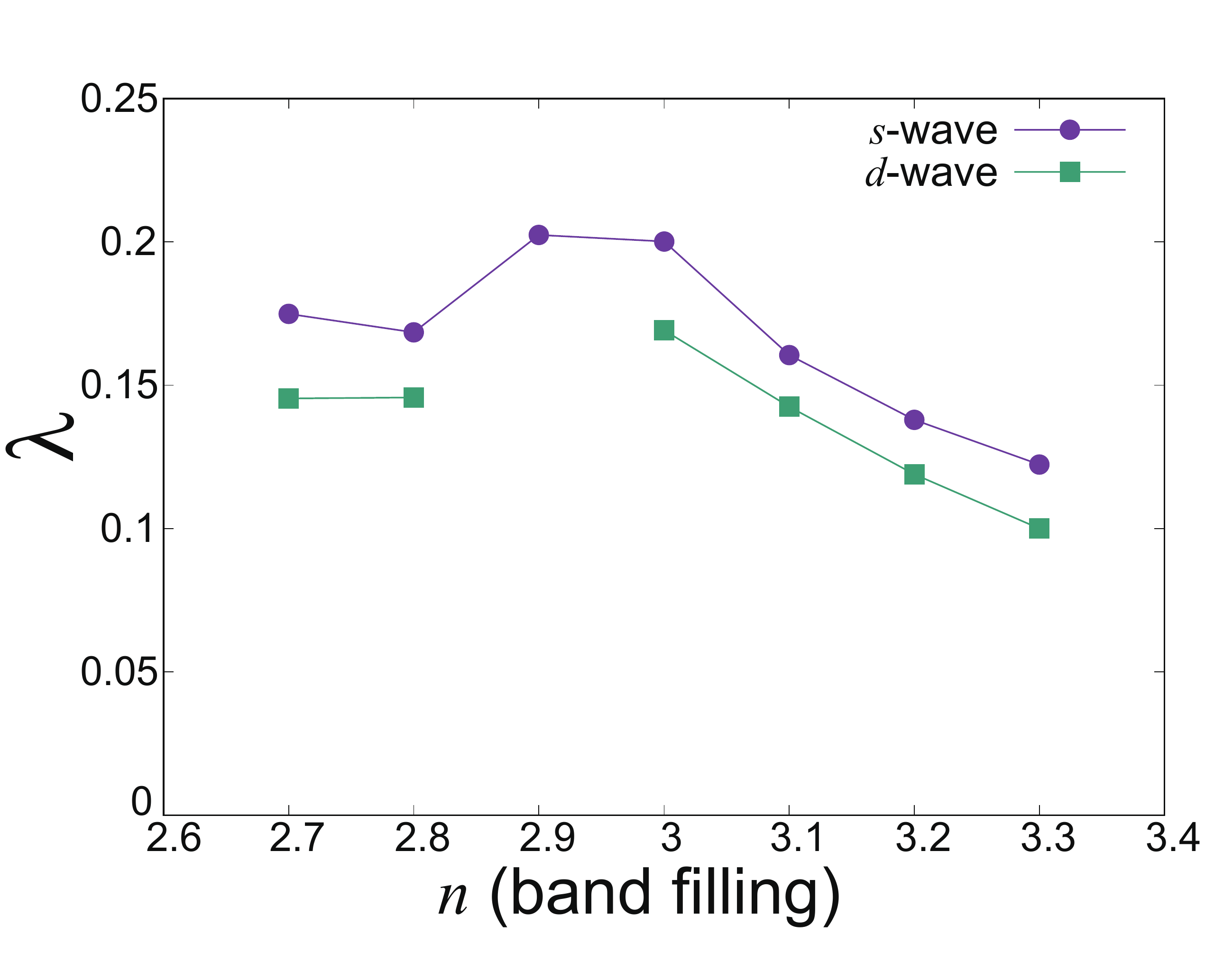}
\caption{Eigenvalue $\lambda$ in the six-orbital model for $s$-wave
and $d$-wave pairings plotted against the band filling. The interaction parameters are $U=2.5\,\mathrm{eV}$, $J=J'=U/10$, and $U'=U-2J$. 
In the six-orbital model, since the Wannier orbitals are localized to each Cu atom, we use $U=2.5$~eV, which is close to values evaluated for the conventional cuprates~\cite{Sa40,Sa41,Sa42}.
}
\end{center}
\label{fig:6orbitallambda}
\end{figure}

\begin{figure}[t]
\centering
\begin{center}
\includegraphics[width=\linewidth]{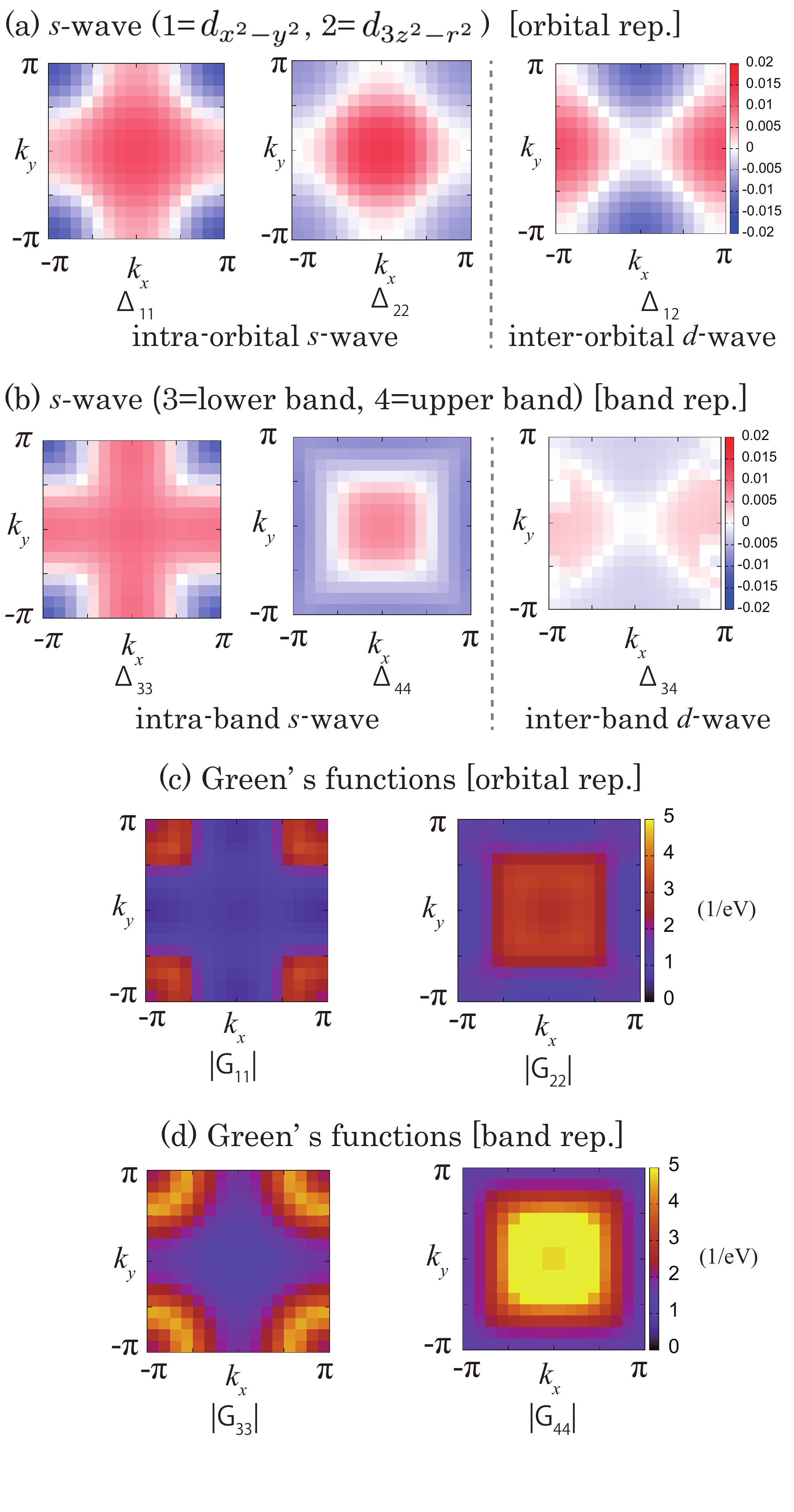}
\caption{The $s$-wave gap function in the six-orbital model in the orbital representation (a) or the band representation (b), along with 
Green's function  the orbital representation (c) 
or the band representation (d). The parameter values adopted are $n=3.0$, $U=2.5\,\mathrm{eV}$, $J=J'=U/10$, and $U'=U-2J$.}
\end{center}
\label{fig:6orbitalgap}
\end{figure}

\subsection{Dependence on the level offset between the two orbitals}
So far we have adopted the tight-binding parameter values (the hoppings and the on-site energies) estimated by first principles band calculation and Wannierization. The obtained maximum value of $\lambda$ is about 0.3 at $T=0.01$\,eV, which is not large enough to explain the $T_c\sim 73$\,K experimentally 
found in $\mathrm{Ba_{2}CuO_{3+\delta}}$~\cite{Ba}. If we stick to these parameter values, some additional pairing mechanisms (e.g., electron-phonon interaction) which boost the $T_c$ would be required. We note, however, that there are some ambiguities in the parameter values in the Hamiltonian, particularly the energy level offset between $d_{x^2-y^2}$ and $d_{3z^2-r^2}$ orbitals. First, in the actual samples used in the experiment with excess oxygens, the holes doped into the planes may attract the apical oxygen more strongly than theoretically estimated here, as we have mentioned above.  
Indeed, the theoretical average apical height, 1.98\,\AA, estimated here for $\mathrm{Ba_{2}CuO_3}$, is larger than the experimental value (1.86\,\AA). The excess oxygens themselves would repel the apical oxygens, but this effect should be insignificant at Cu site 1, where no additional oxygens can be coordinated. The lowered apical oxygens at site 1  would further push up the $d_{3z^2-r^2}$ energy level.  Second, the level offset may be affected by correlation effects that are not taken into account in the first principles calculation.
For instance, Ref.~\cite{LDADMFT} studied a nickelate superlattice system where the $d_{x^2-y^2}$ and $d_{3z^2-r^2}$ levels are inverted, and found that the correlation effect taken into account with the dynamical mean field theory pushes  the $d_{3z^2-r^2}$ band just above the Fermi level to make its Fermi surface vanishing.

With these considerations, let us probe 
how the eigenvalue $\lambda$ against the band filling 
changes when the level offset
\[
\Delta E\equiv E_{d_{3z^{2}-r^{2}}}-E_{d_{x^{2}-y^{2}}}
\]
is varied, in order to seek a possibility for a further enhancement of superconductivity. 
In Figure 13, the eigenvalue of the Eliashberg equation of the two-orbital model is plotted against the band filling for various values of $\Delta E$, 
where the $d_{3z^2-r^2}$ level is changed.  
The result does reveal intriguing features.  
If we first focus on the region around quarter-filling $n=0.5$ (one electron per two orbitals, corresponding to an oxygen content of $3+\delta=3.375$), 
the $\lambda$
is enhanced as $\Delta E$ increases. Here we find that the dominating pairing symmetry changes from $s$-wave to $d$-wave. This is because the system 
approaches a half-filled single-band system for higher $d_{3z^2-r^2}$ level, favoring $d_{x^2-y^2}$-wave pairing~\cite{Dkato}. This resembles the situation in the conventional cuprates with nearly $d^9$ electron configuration (three electrons per two $e_g$ orbitals), where a sufficiently {\it low} $d_{3z^2-r^2}$-level results in an effective single-band system comprising the $d_{x^2-y^2}$ orbital and hence favors $d$-wave pairing~\cite{cu1, cu2, cu3, cu4}: The 
difference is that the $d_{3z^2-r^2}$ is moved away from the main band 
upward in the present case or downward in the latter. However, the eigenvalue $\lambda$ here is not so high as in the typical single-band cuprates such as $\mathrm{HgBa_{2}CuO_4}$ because the band width is narrower. 

If we now turn to the region around $n=1$, $\Delta E$ smaller than the original value somewhat enhances the $s$-wave $\lambda$. We find that a smaller $\Delta E$ increases 
$\Gamma_{1111}$ and $\Gamma_{1122}$ (not shown), and we speculate that the near degeneracy of the two orbitals favors the inter-orbital pairing through the enhancement of $\Gamma_{1122}$ (the intra-orbital scattering 
of inter-orbital pairs). The enhancement of $\lambda$ upon reducing $\Delta E$ further confirms our statement in Sec. III C that the inter-orbital hybridization, which induces  inter-orbital pairing,  is favorable for superconductivity.

What is even more interesting and realistic is the case 
of larger $\Delta E$ in the $n=1$ regime (which does correspond to $3+\delta=3.25$, close to the experimental situation~\cite{Ba}). There, $\lambda$ corresponding to the $s$-wave in Fig.13 is strongly enhanced when $\Delta E$ is increased to some extent from its original value.
%corresponding to $s$-wave is strongly enhanced around $\Delta E$ larger by 0.3-0.5\,eV than the original value. %修正修正%
In this parameter regime, we find that the 
$s$-wave strongly dominates over the $d$-wave. The maximum value of the $s$-wave $\lambda$ is as large as 0.6, which is close to the value obtained for $\mathrm{HgBa_{2}CuO_{4}}$, a superconductor with $T_c\simeq 100$\,K~\cite{Ni}. 
%Hence we may say that the maximum $\lambda$ obtained in this situation is large enough to explain $T_c\simeq 70$\,K observed in $\mathrm{Ba_{2}CuO_{3+\delta}}$.  %70K削除

In fact, for $\Delta E$ where $\lambda$ is optimized, the bottom of the $d_{3z^2-r^2}$ band is close to the Fermi level (as indicated in inset of Fig.~13). Recently, such a band lying just above or below the Fermi level is referred to as an ``incipient band'', and has received attention, especially in the context of the iron-based superconductors~\cite{inc, inc1, inc2, inc3, inc4, inc9, inc5, inc6, inc7, inc8}, where hole bands lying just below the Fermi level are observed in some materials~\cite{Iimura,KFe2Se2,KFe2Se2ARPES,STO,STO2,XJZhou,Takahashi,inc4,inc3,LiOH}. In a wider context, the possibility of the occurrence or strong enhancement of superconductivity due to an incipient band has long 
been proposed for multiband Hubbard models on various types of lattices~\cite{KurokiArita,bi6, Matsumoto2018, KobayashiAoki, Aokireview, Sayyad, Misumi, Mo, Mo2, bi11, bi12, bi13, bi14, bi15, bi16, bi17,twoleg_s}. A salient feature 
in these cases is that the gap function typically exhibits nodeless ``$s\pm$-wave" symmetry. Indeed, the gap function obtained for the present two-orbital model exactly has a nodeless $s\pm$-wave symmetry, as displayed in Fig.~14. Another prominent feature in the gap function is that the inter-orbital pairing (off-diagonal element in the orbital representation) is suppressed compared to the case with the original value of $\Delta E$ (Fig.~6). This may seem to contradict with what we have concluded previously, namely, that the inter-orbital pairing is favorable for the superconductivity. We shall further discuss this issue in Sec. IV B.
Also, 
if we look in Fig.~15 at the effective pairing interactions for the incipient band case, we can find that $\Gamma_{1221}$, which describes the interorbital 
scattering of intraorbital pairs, is strongly enhanced. We shall come back to the relation between the present model and those in the previous studies also in Sec. IV.

\begin{figure}[h]
\centering
\begin{center}
\includegraphics[width=\linewidth]{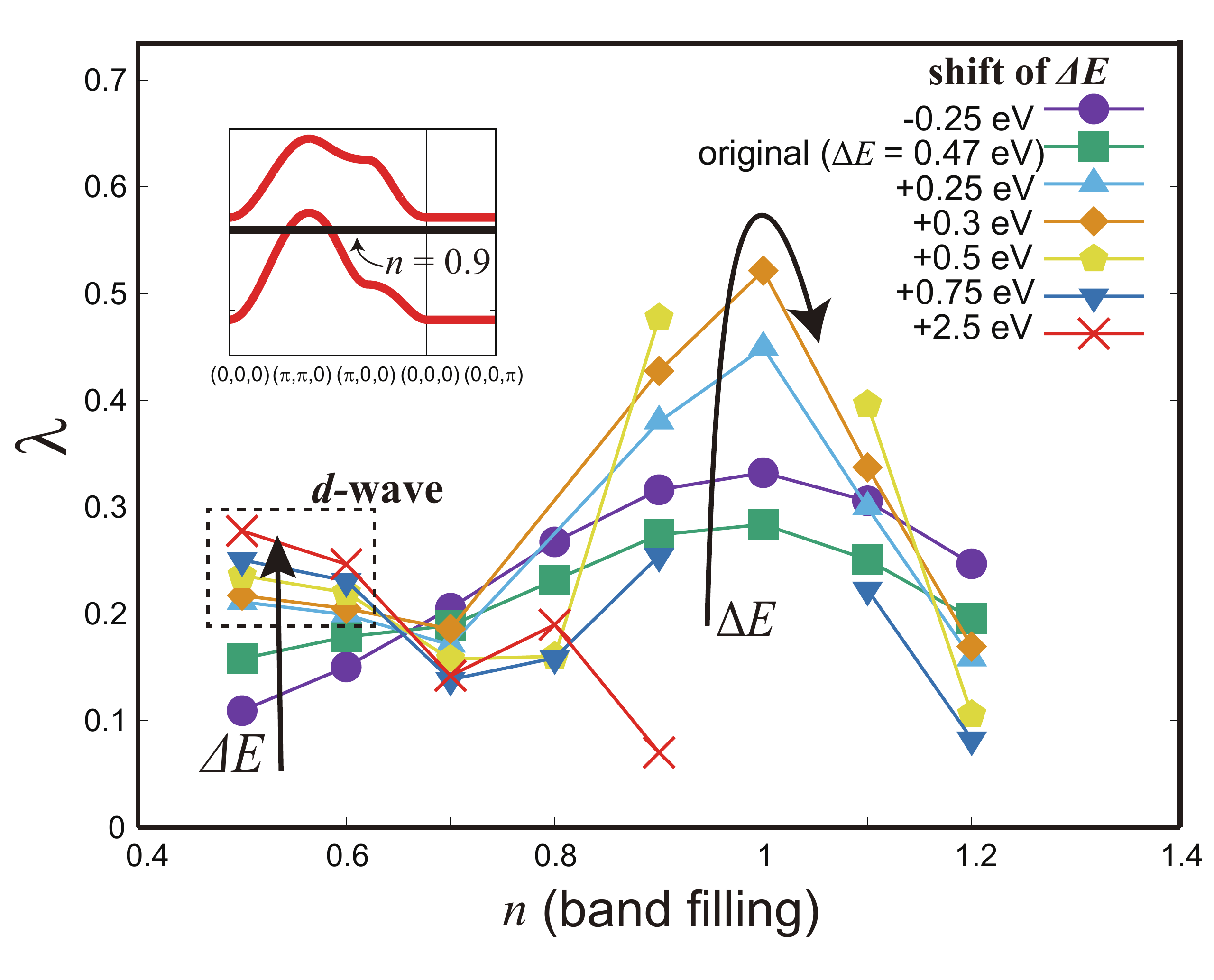}
\caption{The largest eigenvalue $\lambda$ 
of the two-orbital model plotted against the band filling for various values of the level offset, $\Delta E\equiv E_{d_{3z^{2}-r^{2}}}-E_{d_{x^{2}-y^{2}}}$. The original value is $\Delta E=0.47\,\mathrm{eV}$. The pairing symmetry is $s$-wave, except for those symbols marked with a dashed square where the symmetry is $d$-wave. The interaction parameters are $U=2.0\,\mathrm{eV}$, $J=J'=U/10$, and $U'=U-2J$.  The inset depicts the bare band structure with $\Delta E$ increased by +0.5\,eV. The horizontal black line represents the Fermi level for $n=0.9$.
%追加
We note that the FLEX calculation did not converge for some cases around $n=1$ due to large spin fluctuations, and hence the data points for those cases are missing. 
}
\end{center}
\label{fig:onsite}
\end{figure}

\begin{figure}[h]
\centering
\begin{center}
\includegraphics[width=\linewidth]{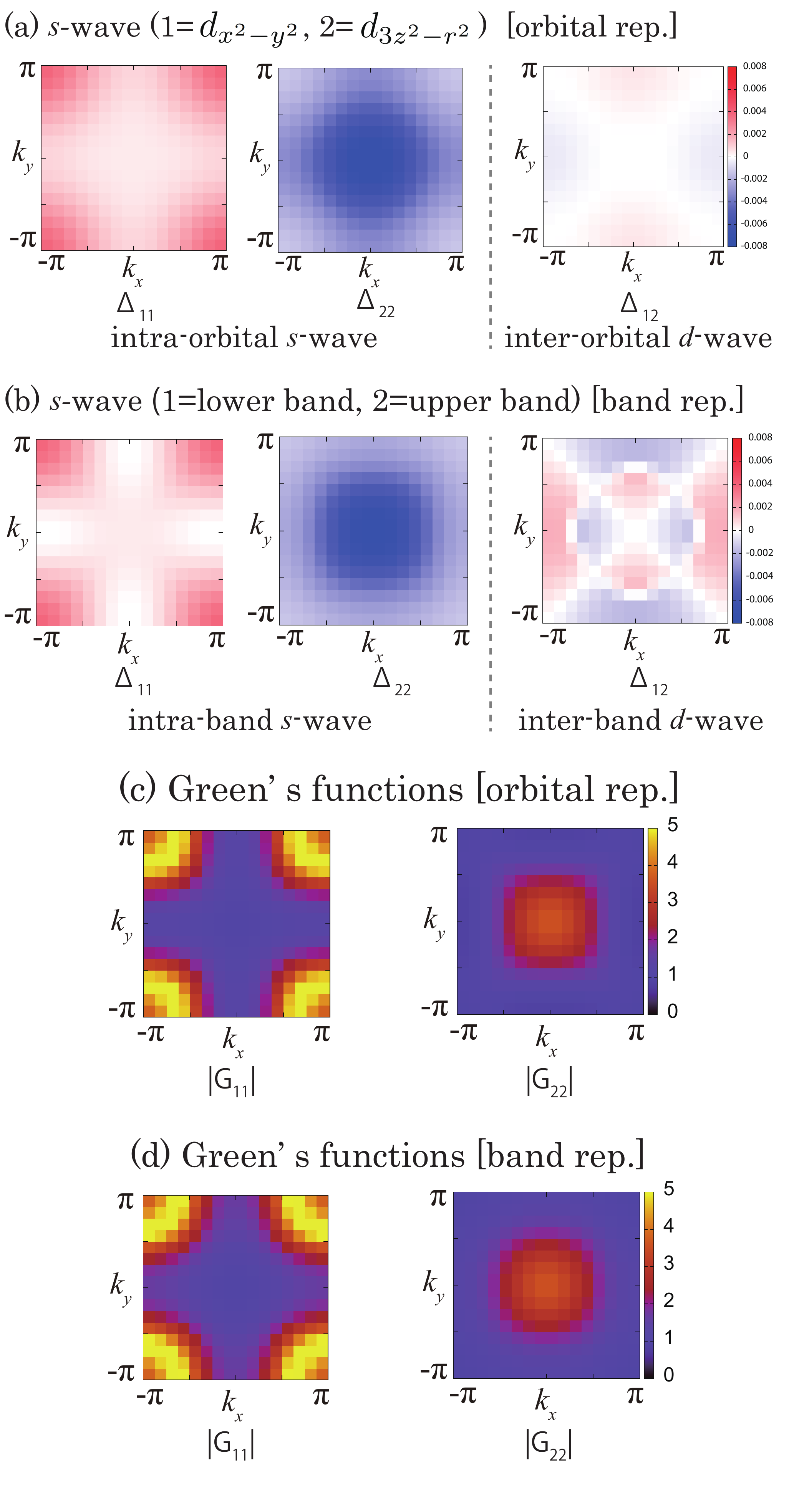}
\caption{The $s$-wave gap function for the two-orbital model in the incipient-band case in the orbital representation (a) and the band representation (b), 
along with Green's function in the orbital representation (c) and the band representation (d). The interaction parameters are $U=2.0\,\mathrm{eV}$, $J=J'=U/10$, $U'=U-2J$, $n=0.9$, and $\Delta E=\Delta E_{\mathrm{original}}+0.5\,\mathrm{eV}$.
%追加
%%
}
\end{center}
\label{fig:incipient_gap}
\end{figure}

\begin{figure}[h!]
\centering
\begin{center}
\includegraphics[width=\linewidth]{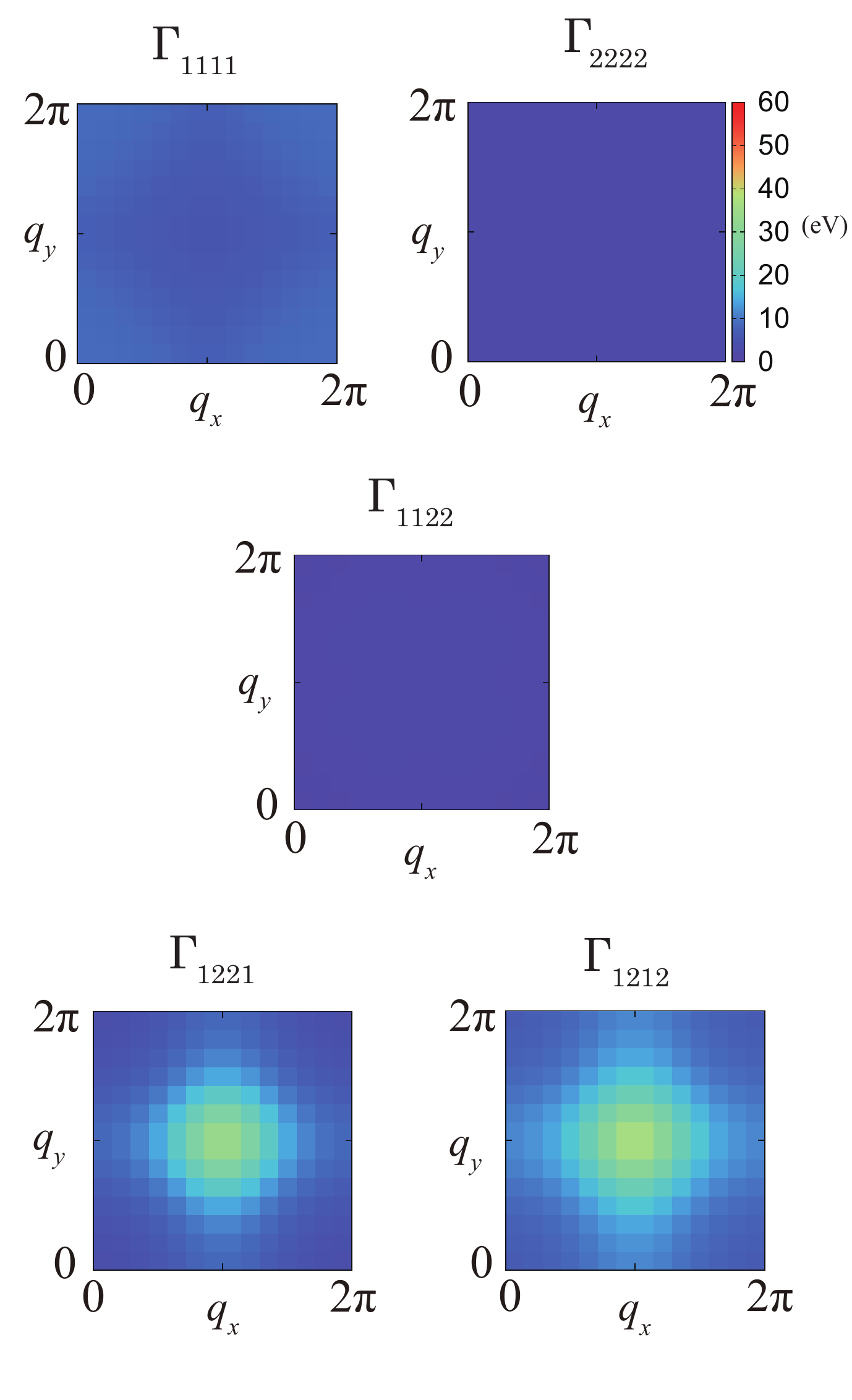}
\caption{The pairing interactions for the two-orbital model in the incipient-band case. The interaction parameters are $U=2.0\,\mathrm{eV}$, $J=J'=U/10$, $U'=U-2J$, $n=0.9$, and $\Delta E=\Delta E_{\mathrm{original}}+0.5\,\mathrm{eV}$.}
\end{center}
\label{fig:incipient_Gamma}
\end{figure}

\section{Discussions}

\subsection{Relation with the K$_{2}$NiF$_{4}$-type structure}
In the present Lieb-lattice model, $s\pm$-wave and $d_{x^2-y^2}$-waves are found to compete with each other. 
A close competition between the $s\pm$ and $d_{x^2-y^2}$ has  been found in a previous theoretical study, where the $\mathrm{K}_{2}\mathrm{NiF}_{4}$-type structure with a reduced apical oxygen height was adopted~\cite{BaRPA,@}. 
It is an intriguing problem how these are possibly 
related.

In Fig.~4(d), we have shown the first principles band structure 
of $\mathrm{Ba}_{2}\mathrm{CuO}_{4}$ 
in the $\mathrm{K}_{2}\mathrm{NiF}_{4}$-type structure. We can 
notice that the band structure of the two-orbital model for the $\mathrm{K}_{2}\mathrm{NiF}_{4}$-type structure is similar to that of 
$\mathrm{Ba}_{2}\mathrm{CuO}_{3}$ in the Lieb-lattice structure [Fig.~4(b)], except for the band width. This is in fact understandable because, in the Lieb lattice, site-2 and site-3 orbitals extend toward site 1 [Fig.~4(c), lower right panel], 
so that they can be regarded as playing a role of the oxygen $2p_\sigma$ orbitals in the $\mathrm{K}_{2}\mathrm{NiF}_{4}$ structure; in this structure, the oxygen $2p$ orbitals have an energy somewhat lower than the Cu $3d$ orbitals, while in the Lieb lattice, the site-2 and site-3 orbitals have somewhat higher energies than the site-1 orbitals, presumably because the apical oxygen height at site-2 and site-3 are lower than at site-1. 
The band width of the latter is narrower than the former because the electron hoppings between Cu site-1 orbital and the Cu site-2 and site-3 orbitals are smaller than those between Cu $3d$ and O $2p$ orbitals. We can therefore state that, starting from the conventional CuO$_2$ plane and removing the oxygens to form a Lieb lattice, we 
unexpectedly encounter an analog of the $\mathrm{CuO_{2}}$ plane, on a smaller energy scale. Since the band structures are similar, so are the FLEX results. In Figs.~16 and 17, we show the FLEX result for the two-orbital model of the K$_2$NiF$_4$-type structure. We again end up with a close competition between $s\pm$-wave and $d_{x^2-y^2}$-wave pairings, which is qualitatively consistent with the previous random-phase approximation study~\cite{BaRPA}. 
%追加
We note that the eigenvalue $\lambda$ for the K$_{2}$NiF$_{4}$-type structure in Fig.~16 is larger than that for the Lieb-lattice-type structure in Fig.~5 because of the wider band  width, which might seem more consistent with the experiment~\cite{Ba} from the viewpoint of the high $T_{c}$, but this, of course, is not the case because the K$_{2}$NiF$_{4}$-type structure does not take into account the large amount of oxygen vacancies observed experimentally.

\begin{figure}[b]
\centering
\begin{center}
\includegraphics[width=\linewidth]{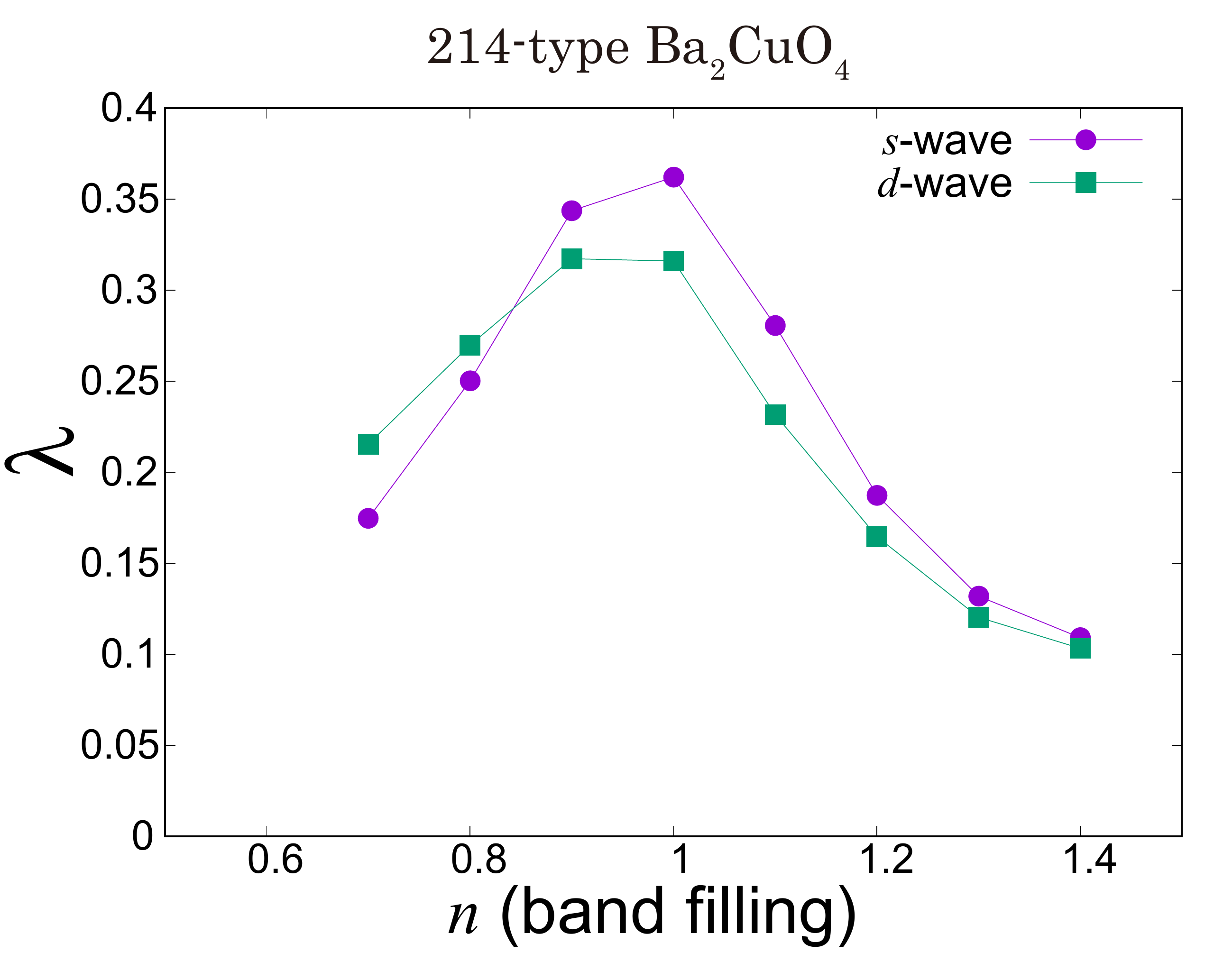}
\caption{Eigenvalue $\lambda$ in the two-orbital model of the K$_2$NiF$_4$-type structure for $s$-wave and $d$-wave pairings plotted against the band filling. The interaction parameters are $U=3.0\,\mathrm{eV}$, $J=J'=U/10$, and $U'=U-2J$. }
\end{center}
\label{fig:214lambda}
\end{figure}

\begin{figure}[t]
\centering
\begin{center}
\includegraphics[width=\linewidth]{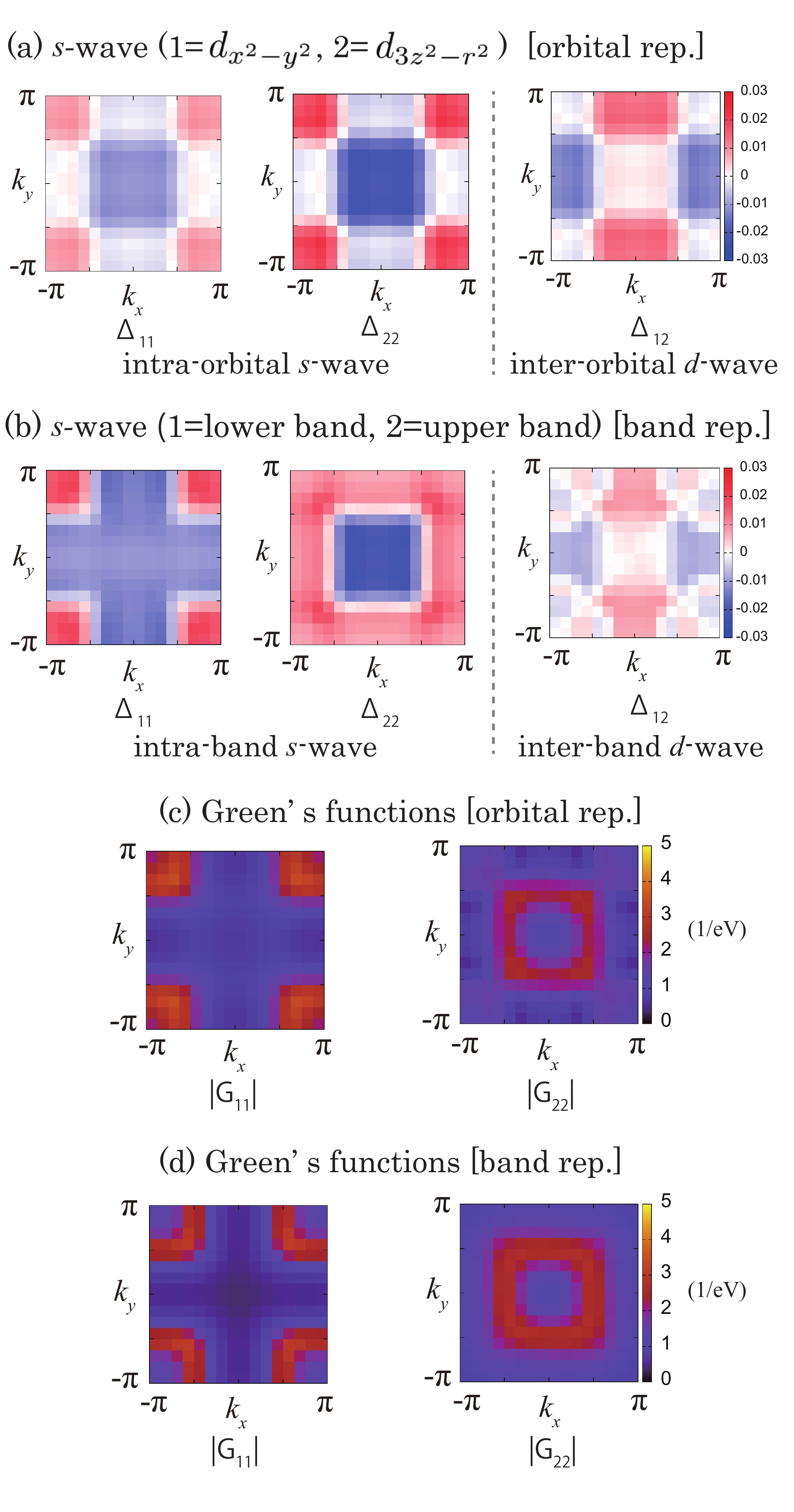}
\caption{The $s$-wave and $d$-wave gap functions in the two-orbital model of the K$_2$NiF$_4$-type structure in the orbital representation (a) or 
band representation (b), along with Green's function in the orbital representation (c) or band representation (d). The parameter values are $n=1.0$, $U=3.0\,\mathrm{eV}$, $J=J'=U/10$, and $U'=U-2J$.}
\end{center}
\label{fig:214gap}
\end{figure}

From the above consideration, we can make an interesting 
observation that the relation between the two-orbital and six-orbital models for the Lieb lattice is analogous to the relation between the single-band Hubbard model and  the three-band $d$-$p$ model in the conventional cuprates.

\subsection{Relation with the bilayer model}
We have found that $s\pm$-wave superconductivity is strongly enhanced in the Lieb-lattice model when the $d_{3z^2-r^2}$ band is raised so that it becomes incipient. Such a strong enhancement of $s\pm$-wave superconductivity reminds us of the bilayer Hubbard model on a square lattice~\cite{bi, bi2, bi3, bi4, bi5, bi6, bi7, bi8, bi9, bi10, bi11, bi12, bi13, bi14, bi15, bi16, bi17}, where 
enhanced superconductivity is found when one of the bands is incipient~\cite{bi6, bi11, bi12, bi13, bi14, bi15, bi16, bi17}. However, the bilayer Hubbard model is a single-orbital (one orbital per site) system with two sites per unit cell. The enhancement of superconductivity in such a system is mediated by spin fluctuations originating from the on-site $U$, which gives rise to the pairing interaction between the bonding and antibonding bands, both of which have equal weight of the two orbitals in a unit cell.
Similar enhancement mechanism of superconductivity in multiorbital systems 
has been discussed in the context of the iron-based superconductors, where portions of the electron and hole Fermi surfaces having the same orbital character interact via the effective interaction ($\Gamma_{llll}$ in the present notation) enhanced mainly by the intra-orbital $U$~\cite{ironkk}. In the present Lieb-lattice model, and also in the two-orbital model of the K$_2$NiF$_4$-type structure, the situation is distinct in that the orbital character is quite different 
between the two bands. The lower band has a strong $d_{x^2-y^2}$ character, while the upper band is dominated by $d_{3z^2-r^2}$ character, so that 
here the inter-orbital interactions $(U',J,J')$  should be the key.
The present view that inter-orbital interactions play an important role can be reinforced  from the pairing interactions presented in Fig.~15, where the inter-orbital pair scattering 
vertices ($\Gamma_{1221}$ and $\Gamma_{1212}$) are large, and also from the gap functions in Fig.~14, where the sign of the gap function is reversed between $d_{x^{2}-y^{2}}$ and $d_{3z^{2}-r^{2}}$ orbitals. Figure 14 
also shows that the orbital and the band representations 
resemble with each other, which implies the two bands have different orbital characters.

Can we identify the reason why superconductivity is enhanced even when the incipient band has an orbital character different from that of the main band?  
Let us propose a succinct way to understand this. As shown in Appendix C, the single-orbital (with one orbital per site) Hubbard model on a bilayer square lattice [Fig.~18(a)] with an on-site interaction $U$ can be transformed into a two-orbital Hubbard model on a (monolayer) square lattice with all the on-site intra- and inter-orbital interactions being $U/2$~\cite{ShinaokaTransfm}. In this transformation, the bonding and antibonding orbitals in the bilayer system, comprising the two sites connected by the vertical interlayer hopping $t_\perp$, translate to the $d_{x^2-y^2}$ and $d_{3z^2-r^2}$ orbitals in the monolayer system, 
with $\Delta E$ playing a role of $2t_\perp$. In the bilayer Hubbard model, whose noninteracting band structure is depicted in Fig.~18(b), superconductivity is found to be strongly enhanced when one of the bands becomes (nearly) incipient upon increasing $t_\perp$~\cite{bi6, bi11, bi12, bi13, bi14, bi15, bi16}.

From the viewpoint of the transformation introduced here, the result for the two-orbital model of the Lieb lattice thus corresponds to that of the bilayer model in that superconductivity is strongly enhanced when the upper band becomes incipient upon increasing $\Delta E$. The gap function of the Lieb two-orbital model also resembles that obtained for the bilayer model. We show in Fig.~18(c) the gap function of the bilayer Hubbard model in the band representation. (Note that the band representation in the bilayer model corresponds to the orbital representation in the two-orbital model.) The parameter values are determined from those for the Lieb two-orbital model with $\Delta E$ increased by +0.5\,eV 
 using the transformation given in Appendix C. The nodeless $s\pm$-wave gap function indeed resembles that of the two-orbital model.  By ``nodeless,'' we mean that the gap 
within each band does not change sign, not only on the Fermi surface but 
over the entire Brillouin zone. Also, the enhancement of the interaction $\Gamma_{1221}$ for the inter-orbital scattering of intra-orbital pairs 
(Fig.~15) corresponds to the dominant pairing interaction in the bilayer model that induces inter-band scattering of intra-band pairs. All these resemblances between the two-orbital model and the bilayer model suggest that the transformation is approximately valid even when $U\neq U'\neq J$. On the other hand, if we look more closely, the gap function of the bilayer model is nearly constant within each band [Fig.~18(c)], whereas that of the present model exhibits momentum dependence, which roughly has a 
 $\cos(k_x)+\cos(k_y)+{\rm const.}$ form. Namely, in the bilayer model, the pairing in real space occurs basically within the same unit cell (connected by $t_\perp$), while in the two-orbital model a mixing of intra- and inter-unit cell pairings takes place.

%追加修正
The interorbital repulsion $U'$ is usually known to enhance charge or orbital fluctuations~\cite{Takimoto}, which generally compete with spin fluctuations in mediating Cooper pairing because the charge and orbital (spin) fluctuations give rise to attractive (repulsive) pairing interactions. An intriguing point to note in the present case is that $U'$ plays a crucial role in enhancing spin fluctuations, as can be captured from the analogy with the bilayer Hubbard model where spin fluctuations solely dominate.

The two-orbital to bilayer transformation appears to be approximately valid only when $ \Delta E$ is not too small; namely, $s$-wave dominates over $d$-wave 
when $\Delta E$ is small in the present two-orbital model, while $d$-wave is dominant in the bilayer model with small $t_\perp$~\cite{bi11}. This is presumably because  the $\cos (k_x)-\cos(k_y)$  form of the hybridization cannot be transformed into the bilayer square lattice form of the Hamiltonian (with 
the off-diagonal elements in  Eq.~(C5) in Appendix C having tetragonal symmetry), so that the  transformation loses its validity as $\Delta E$ becomes smaller 
than the inter-orbital hopping between $d_{x^2-y^2}$  and $d_{3z^2-r^2}$ orbitals. When $\Delta E$ is large, on the other hand, the inter-orbital hybridization loses its significance, so that the transformation becomes more valid. However, when $\Delta E$ is small, the effect of the hybridization will be prominent, so that $d$-wave gives way to $s$-wave, in contrast to the bilayer model case, because, as mentioned in Sec. III C, the inter-orbital hopping makes the $s$-wave pairing more favorable.

We mentioned in Secs. III C and III D that the inter-orbital hybridization is favorable for superconductivity. However, we find that this is not the case when $\Delta E$ is large as in the incipient band situation; there, if we turn off the inter-orbital hopping $t_{12}$, the inter-orbital component of the gap function $\Delta_{12}$ vanishes, but $\lambda$ is {\it enhanced}, namely, the better correspondence to the bilayer model is more favorable for superconductivity. 
%追加
Since the inter-orbital and the incipient-band enhanced pairing mechanisms are essentially different, whether the presence of inter-orbital pairing is favorable for superconductivity or not depends on the magnitude of the level offset between the two orbitals.

We note that apart from the problem of $\mathrm{Ba_{2}CuO_{3+\delta}}$, the occurrence of $s\pm$-wave superconductivity in the cuprates was also proposed for a model~\cite{CuO2mltheory} of highly overdoped CuO$_2$ monolayer grown on $\mathrm{Bi}_{2}\mathrm{Sr}_{2}\mathrm{CaCu}_{2}\mathrm{O}_{8+\delta}$~\cite{CuO2mlexp}. In this model, the $d_{3z^2-r^2}$ band lies below the $d_{x^2-y^2}$ band as in the conventional cuprates, but due to the large amount of holes, the Fermi level not only intersects the $d_{x^2-y^2}$ band, but also intersects the top of $d_{3z^2-r^2}$ band. This resembles the incipient-band situation of the present two-orbital model, if we make an electron-hole transformation.

\begin{figure}[t]
\centering
\begin{center}
\includegraphics[width=\linewidth]{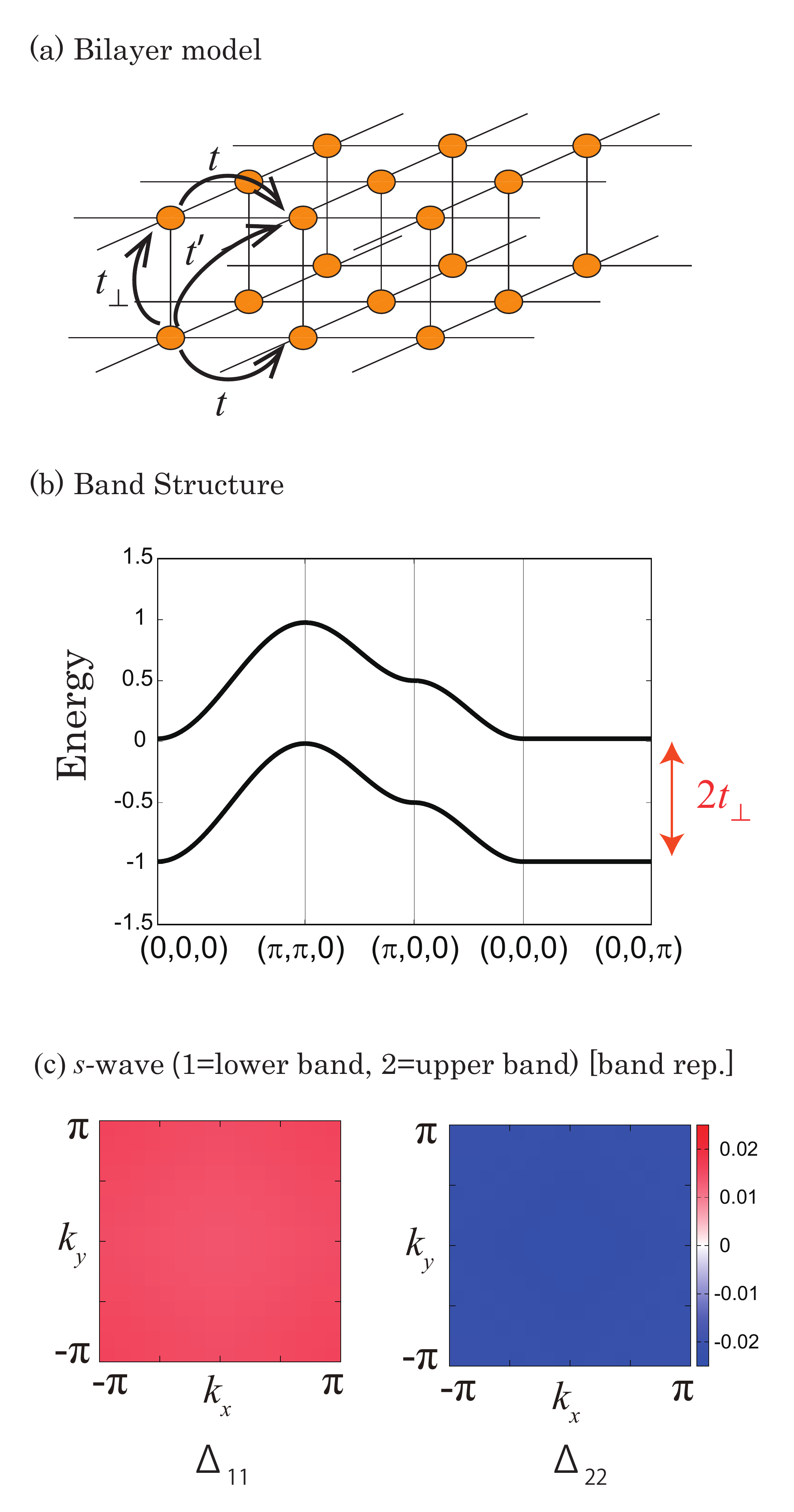}
\caption{(a) The bilayer square lattice, (b) band structure of this model, and (c) the $s$-wave gap functions of the bilayer Hubbard model in the band representation. The parameter values are $n=0.9$, $t=-0.12$, $t'=0.01$, $t_{\perp}=0.5$, and $U=3$.}
\end{center}
\label{fig:bilayer}
\end{figure}

%$U'$, $J$ dependence.
%
%comparison of the gap function between Lieb two-orbital, 214 two-orbital and bilayer for the original modeld and also at the incipient point.

\section{CONCLUSIONS}

In the present study, we have proposed the Lieb lattice as a candidate for the lattice structure of the newly discovered superconductor 
$\mathrm{Ba_{2}CuO_{3+\delta}}$. We have shown from the total energy that 
the proposed lattice structure is almost as stable as 
the chain-type structure that is known to exist. The dynamical stability of the proposed structure is also shown through a phonon calculation. 

Applying a FLEX approximation to the relevant two-orbital and six-orbital models derived from the first-principles band calculation, 
we find that {\it coexistence of intra- and inter-orbital pairings} arises due to the relatively small energy level offset between the $d_{x^2-y^2}$ and $d_{3z^2-r^2}$ orbitals. As for the pairing symmetry, 
$s$-wave and $d$-wave pairings closely compete with each other, with 
the former dominating. Superconductivity is optimized around the band filling corresponding to the oxygen content of $3+\delta=3.25$, which is close to that of the actual material. 

The maximum eigenvalue of the Eliashberg equation 
is not large enough to explain the observed $T_{c}$.  
While a cooperation with other pairing glues such as phonons may be necessary to fully understannd the experiment, 
we have proposed an alternative scenario for explaining the observed $T_{c}$ by varying the level offset between the two orbitals, which is motivated from the consideration that the level offset may be larger in the actual material Ba$_{2}$CuO$_{3+\delta}$ than its first principles estimation for Ba$_{2}$CuO$_{3}$.
We have indeed found that $s\pm$-wave superconductivity is strongly enhanced when the $d_{3z^2-r^2}$ band is raised in energy so that it becomes nearly 
{\it incipient} around the band filling corresponding to the oxygen content of $3+\delta= 3.25$. In this situation, in contrast to the case with smaller level offset, the amplitude of the inter-orbital pairing is small, while the {\it inter-orbital pair scattering} plays an essential role.

We have then noted that both the band structure and the FLEX results resemble those of the two-orbital model for the $\mathrm{K}_{2}\mathrm{NiF}_{4}$-type structure.  We have traced its origin back to the fact that 
the Cu orbitals at sites 2 and 3 in the Lieb lattice play the role of the oxygen $2p_\sigma$ orbitals in the $\mathrm{K}_{2}\mathrm{NiF}_{4}$ structure, so that the electronic structure of the former is analogous to that of the latter.

From this observation, we have further pointed out a relation between the two-orbital model for the Lieb lattice and the Hubbard model on the bilayer square lattice. When one of the bands is incipient, the two models exhibit similar results regarding the enhancement of superconductivity and the nodeless form of the gap function. The resemblance suggests that the transformation between the two-orbital model and the bilayer model, which is shown to be rigorous when the intra- and inter-orbital interactions are equal, is valid to some extent even when the interactions are not equal.

In the present study, we have focused on the 2-1-3 composition, and varied the band filling assuming a rigid band. It will be an interesting and important 
future problem to explicitly investigate the effect of the ``$+\delta$'' excess oxygens.

\begin{acknowledgements}
K.Y., M.O., and K.K. acknowledge valuable discussions
with Daichi Kato. H.A. thanks Core Research for Evolutional
Science and Technology “Topology” project from Japan Science and Technology Agency. This study is supported by
Japan Society for the Promotion of Science KAKENHI Grant
No. JP18H01860.
\end{acknowledgements}

\appendix

\def\thesection{\Alph{section}}

\section{The crystal structures of the Lieb-lattice-type and the chain-type Ba$_2$CuO$_3$}

In the main text we have proposed a Lieb-lattice-type structure for Ba$_{2}$CuO$_3$ and discussed the stability for comparison with the chain-type structure. For their stacking geometry, we show the detail in Fig.~19 
as the side views of these structures that we actually use in the first-principles calculation. For the Lieb lattice, layers are stacked in such a way that the in-plane components of the translation vector between adjacent layers are always $\sim(\frac{1}{4},\frac{1}{4})$ in units of the lattice constants, as depicted by a side view [Fig.~19(a)] and also by a top view (Fig.~20) of the Cu-O planes. This stacking pattern breaks tetragonal symmetry as can be seen from Fig.~20; in fact, one way to strictly preserve this symmetry is to have four layers in a unit cell, where the in-plane components of the translation vector between neighboring layers are $\sim(\frac{1}{4},\frac{1}{4})$, $(\frac{1}{4},-\frac{1}{4})$, $(-\frac{1}{4},-\frac{1}{4})$, and $(-\frac{1}{4},\frac{1}{4})$, but that would result in a very large unit cell. We adopt the structure depicted in Figs.~19 and ~20 to reduce the size of the unit cell and hence the calculation cost.
In practice, however, 
we find that the structure we adopt approximately preserves tetragonal symmetry in that the tight-binding parameters of the obtained models possess tetragonal symmetry within the accuracy $\sim 10^{-4}$\,eV.  
As explained in Appendix B below, these small parameters are disregarded 
in the FLEX calculation, so that the Hamiltonian used in FLEX preserves 
the tetragonal symmetry.

\begin{figure}[h!]
\centering
\begin{center}
\includegraphics[width=\linewidth]{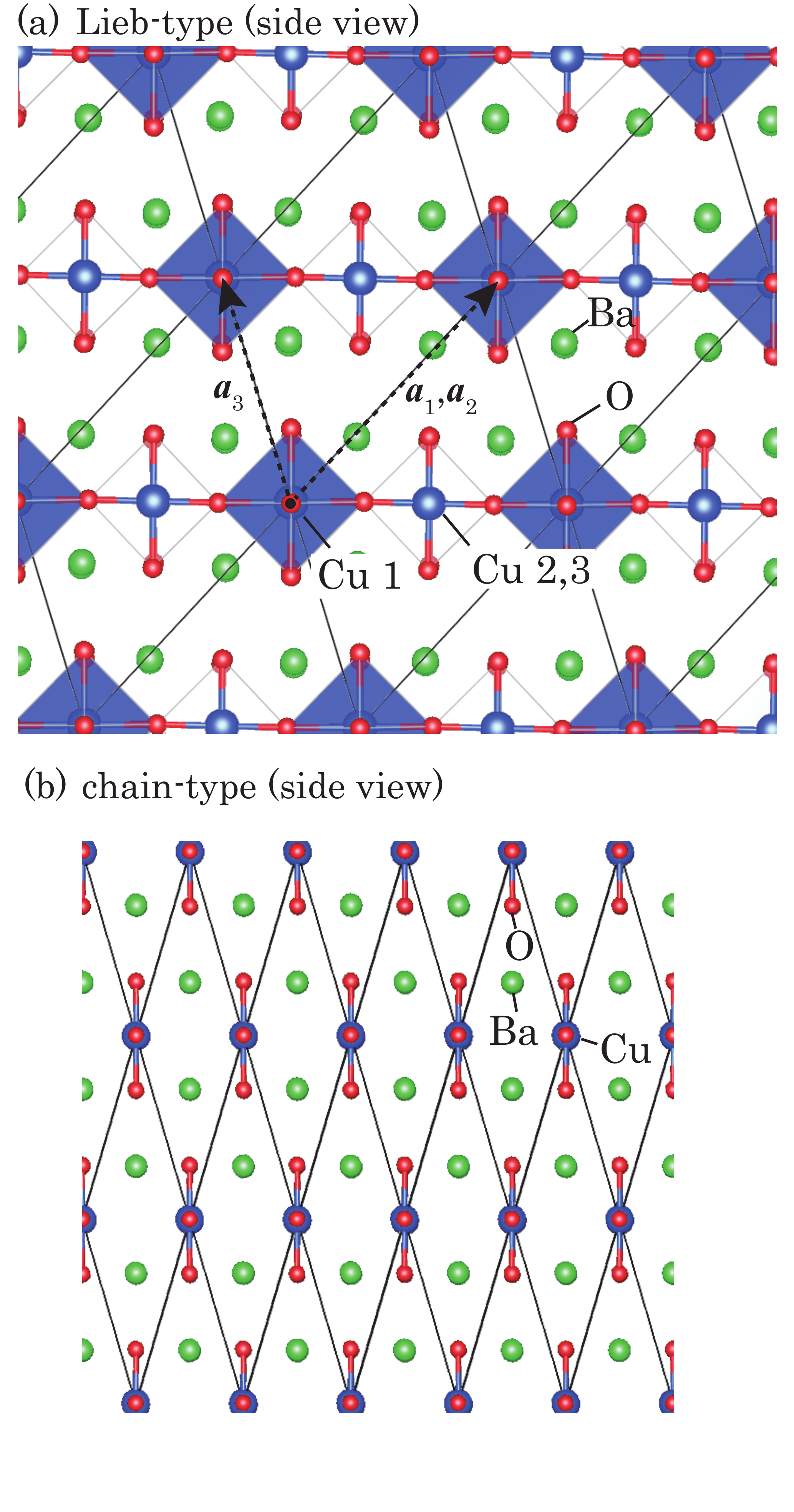}
\caption{Side views of (a) the Lieb-lattice-type and (b) chain-type 
structures, depicted with VESTA software. In panel (a), we indicate the translation vectors $\bm{a}_{1}$, $\bm{a}_{2}$, and $\bm{a}_{3}$. Parallelograms formed by blue lines delineate the unit cells. 
}
\end{center}
\label{fig:Lieb_chain_struct}
\end{figure}

\begin{figure}[h!]
\centering
\begin{center}
\includegraphics[width=\linewidth]{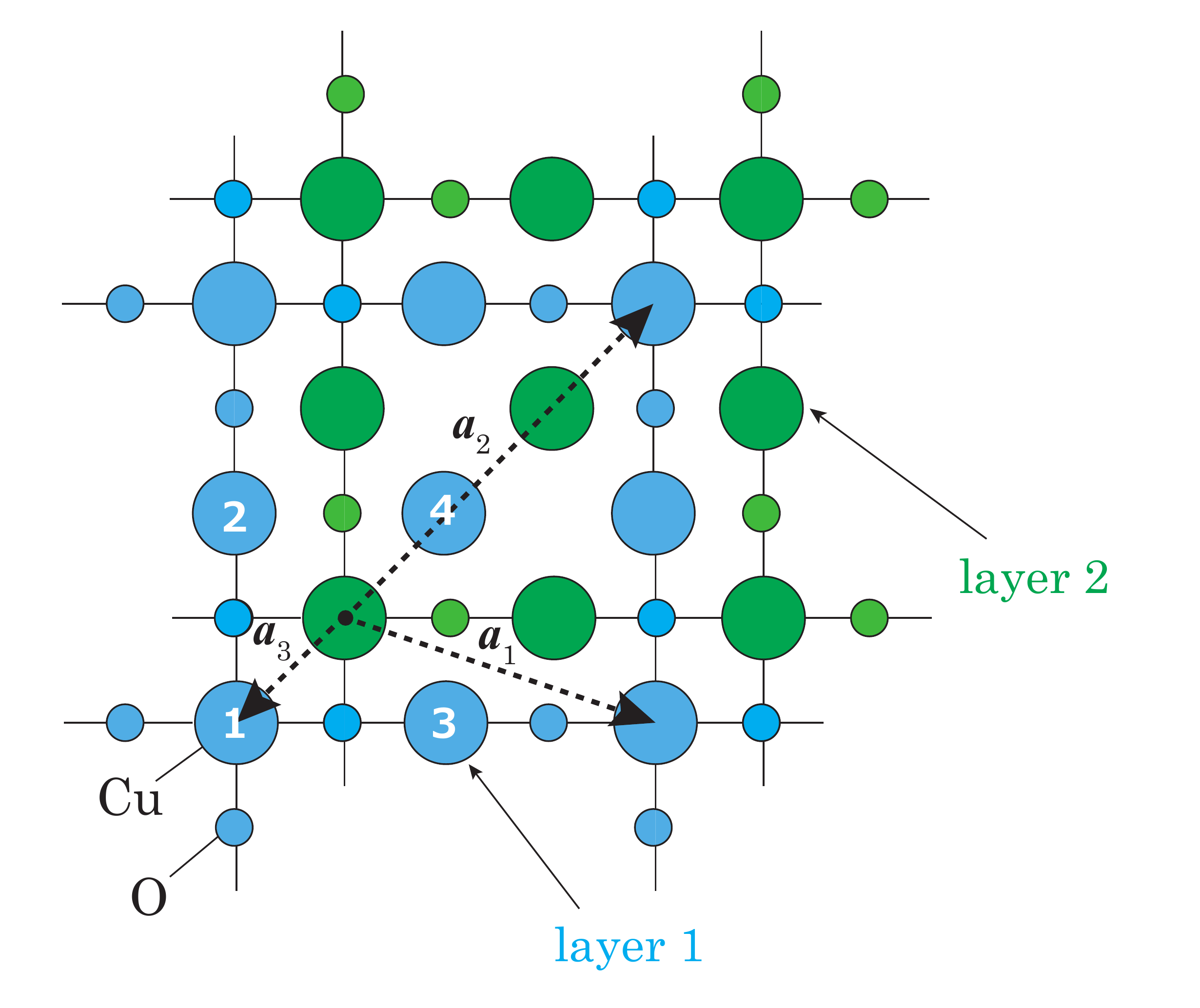}
\caption{A top view of the stacked Cu-O layers of the Lieb-lattice-type structure.  The translation vectors are also indicated as in the previous figure.}
\end{center}
\label{fig:Lieb_stack}
\end{figure}

\section{The tight-binding parameter values for the Lieb lattice models}

Here we give the tightbinding parameters of the six-orbital [Fig.~4(a)] and two-orbital [Fig.~4(b)] Lieb lattice models obtained by the first-principles calculation with WANNIER90. In order to simplify the multiorbital models, we disregard small hopping parameters $|t|<1.0\times10^{-2}$\,eV in the six-orbital model 
or $|t|<5.0\times10^{-2}$\,eV  in the two-orbital model. 

In the six-orbital Lieb lattice model, we label the Wannier orbitals 
in a unit cell are as
\begin{equation}
\begin{pmatrix}
    1
\\ 2
\\ 3
\\ 4 
\\ 5
\\ 6 
\end{pmatrix}=\begin{pmatrix}
    \text{orbital 1} \text{ at site 1}
\\ \text{orbital 2} \text{ at site 1}
\\ \text{orbital 3} \text{ at site 2} 
\\ \text{orbital 4} \text{ at site 2}
\\ \text{orbital 5} \text{ at site 3}
\\ \text{orbital 6} \text{ at site 3} 
\end{pmatrix},
\end{equation}
where the orbital numbers are depicted in Fig.~4(c).
The kinetic energy in the multiorbital Hamiltonian is given 
in terms of the hopping parameters $t_{lm}$ obtained by WANNIER90 as
\begin{equation}
H_{\mathrm{Kin}}=\sum_{\bm{k}lm\sigma}\sum_{\bm{R}}t_{lm}(\bm{R})e^{i\bm{k}\cdot\bm{R}}c^{\dagger}_{\bm{k}l\sigma}c_{\bm{k}m\sigma},
\end{equation}
where $l$ and $m$ denote the orbitals and $\bm{R}$ is the basic translation vector.  
Let us display the hopping integral matrix $t_{lm}(\bm{R})$ [eV] for 
each basic translation vector $\bm{R}$:\\
For $\bm{R}=(0,0,0)$,
\begin{equation}
\begin{pmatrix}
    2.51 & 0 & 0.15 & 0.34 & -0.15 & -0.34
\\ 0 & 2.98 & 0.08 & 0.23 & 0.08 & 0.23
\\ 0.15 & 0.08 & 1.10 & 1.34 & 0 & 0
\\ 0.34 & 0.23 & 1.34 & 2.51 & 0 & -0.04
\\ -0.15 & 0.08 & 0 & 0 & 1.10 & 1.34
\\ -0.34 & 0.23 & 0 & -0.04 & 1.34 & 2.51
\end{pmatrix}.
\end{equation}
For $\bm{R}=(0,-1,0)$,
\begin{equation}
\begin{pmatrix}
    -0.09 & 0.05 & 0 & 0 & 0 & 0
\\ 0.05 & -0.03 & 0 & 0 & 0 & 0
\\ 0.15 & 0.08 & 0 & 0 & 0 & 0
\\ 0.34 & 0.23 & 0 & -0.03 & 0 & -0.04
\\ 0 & 0 & 0 & 0 & 0 & 0
\\ 0 & 0 & 0 & 0 & 0 & 0
\end{pmatrix}.
\end{equation}
For $\bm{R}=(0,+1,0)$,
\begin{equation}
\begin{pmatrix}
    -0.09 & 0.05 & 0.15 & 0.34 & 0 & 0
\\ 0.05 & -0.03 & 0.08 & 0.23 & 0 & 0
\\ 0 & 0 & 0 & 0 & 0 & 0
\\ 0 & 0 & 0 & -0.03 & 0 & 0
\\ 0 & 0 & 0 & 0 & 0 & 0
\\ 0 & 0 & 0 & -0.04 & 0 & 0
\end{pmatrix}.
\end{equation}
For $\bm{R}=(-1,0,0)$,
\begin{equation}
\begin{pmatrix}
    -0.09 & -0.05 & 0 & 0 & 0 & 0
\\ -0.05 & -0.03 & 0 & 0 & 0 & 0
\\ 0 & 0 & 0 & 0 & 0 & 0
\\ 0 & 0 & 0 & 0 & 0 & 0
\\ -0.15 & 0.08 & 0 & 0 & 0 & 0
\\ -0.34 & 0.23 & 0 & -0.04 & 0 & -0.03
\end{pmatrix}.
\end{equation}
For $\bm{R}=(+1,0,0)$,
\begin{equation}
\begin{pmatrix}
    -0.09 & -0.05 & 0 & 0 & -0.15 & -0.34
\\ -0.05 & -0.03 & 0 & 0 & 0.08 & 0.23
\\ 0 & 0 & 0 & 0 & 0 & 0
\\ 0 & 0 & 0 & 0 & 0 & -0.04
\\ 0 & 0 & 0 & 0 & 0 & 0
\\ 0 & 0 & 0 & 0 & 0 & -0.03
\end{pmatrix}.
\end{equation}
For $\bm{R}=(+1,-1,0)$,
\begin{equation}
\begin{pmatrix}
    0 & 0 & 0 & 0 & 0 & 0
\\ 0 & 0 & 0 & 0 & 0 & 0
\\ 0 & 0 & 0 & 0 & 0 & 0
\\ 0 & 0 & 0 & 0 & 0 & -0.04
\\ 0 & 0 & 0 & 0 & 0 & 0
\\ 0 & 0 & 0 & 0 & 0 & 0
\end{pmatrix}.
\end{equation}
For $\bm{R}=(-1,+1,0)$,
\begin{equation}
\begin{pmatrix}
    0 & 0 & 0 & 0 & 0 & 0
\\ 0 & 0 & 0 & 0 & 0 & 0
\\ 0 & 0 & 0 & 0 & 0 & 0
\\ 0 & 0 & 0 & 0 & 0 & 0
\\ 0 & 0 & 0 & 0 & 0 & 0
\\ 0 & 0 & 0 & -0.04 & 0 & 0
\end{pmatrix}.
\end{equation}

In the two-orbital Lieb lattice model, we label the Wannier orbitals as
\begin{equation}
\begin{pmatrix}
    1
\\ 2
\end{pmatrix}=\begin{pmatrix}
    d_{x^{2}-y^{2}} \text{ at site 1}
\\ d_{3z^{2}-r^{2}} \text{ at site 1}
\end{pmatrix}.
\end{equation}
The hopping integral matrix for each $\bm{R}$ is given as follows:\\
For $\bm{R}=(0,0,0)$,
\begin{equation}
\begin{pmatrix}
2.20 & 0 \\ 0 & 2.67 
\end{pmatrix}.
\end{equation}
For $\bm{R}=(\pm1,0,0)$,
\begin{equation}
\begin{pmatrix}
-0.15 & 0.12 \\ 0.12 & -0.11 
\end{pmatrix}.
\end{equation}
For $\bm{R}=(0,\pm1,0)$,
\begin{equation}
\begin{pmatrix}
-0.15 & -0.12 \\ -0.12 & -0.11 
\end{pmatrix}.
\end{equation}

\section{The relation between the two-orbital model and the bilayer model}

Let us explain here the relation between the single-orbital (one orbital per site) bilayer model and the two-orbital model~\cite{ShinaokaTransfm}. 
We label the two sites in a unit cell of the bilayer lattice as $i=1,2$, and the orbitals in the two-orbital model as $a,b=\alpha,\beta$. The on-site interaction part of the Hamiltonian of the bilayer Hubbard model is 

\begin{equation}
H_{\mathrm{int}}^{\mathrm{bilayer}}=U\sum_{m}\sum_{i=1,2}n_{mi\uparrow}n_{mi\downarrow},
\label{bilayer:eq}
\end{equation}
where $m$ labels the unit cell. With a transformation $R$ defined as

\begin{equation}
\begin{split}
 \begin{pmatrix} d_{m\alpha\sigma} \\ d_{m\beta\sigma} \end{pmatrix} = R\begin{pmatrix} c_{m1\sigma} \\ c_{m2\sigma} \end{pmatrix} = \frac{1}{\sqrt{2}}\begin{pmatrix} 1 & 1 \\ -1 & 1 \end{pmatrix} \begin{pmatrix} c_{m1\sigma} \\ c_{m2\sigma} \end{pmatrix}
\end{split},
\end{equation}
we go from the site basis ($c$) to the bonding/antibonding orbital basis ($d$).
Then Eq.(\ref{bilayer:eq}) is cast into 

\begin{equation}
\begin{split}
&\frac{U}{2}\sum_{m,a}n_{ma\uparrow}n_{ma\downarrow}+\frac{U}{2}\sum_{m,a\neq b}n_{ma\uparrow}n_{mb\downarrow}\\
&-\frac{U}{2}\sum_{m,a\neq b}d^{\dagger}_{ma\uparrow}d_{ma\downarrow}d^{\dagger}_{mb\downarrow}d_{mb\uparrow}\\
&+\frac{U}{2}\sum_{m,a\neq b}d^{\dagger}_{ma\uparrow}d^{\dagger}_{ma\downarrow}d_{mb\downarrow}d_{mb\uparrow}.
\end{split}
\end{equation}
Namely, we end up with a two-orbital model where 
the on-site intra- and inter-orbital interactions all 
have the same strength, $U/2$. 

We can also show how the kinetic energy part of the Hamiltonian is transformed. Let $c_{\bm{k}i\sigma}$, $c^{\dagger}_{\bm{k}i\sigma}$ be the Fourier transform of $c_{mi\sigma}$, $c^{\dagger}_{mi\sigma}$. The kinetic energy part is then 
given in momentum space as 

\begin{equation}
\begin{split}
H_{\mathrm{Kin}}^{\mathrm{bilayer}}&=\begin{pmatrix} c^{\dagger}_{\bm{k}1\sigma} & c^{\dagger}_{\bm{k}2\sigma} \end{pmatrix}\hat{H}(\bm{k})\begin{pmatrix} c_{\bm{k}1\sigma} \\ c_{\bm{k}2\sigma} \end{pmatrix}\\
&=\begin{pmatrix} c^{\dagger}_{\bm{k}1\sigma} & c^{\dagger}_{\bm{k}2\sigma} \end{pmatrix} \begin{pmatrix} 
\varepsilon_{1}(\bm{k}) & \varepsilon'(\bm{k}) \\ \varepsilon'(\bm{k}) & \varepsilon_{2}(\bm{k}) \end{pmatrix} \begin{pmatrix} c_{\bm{k}1\sigma} \\ c_{\bm{k}2\sigma} \end{pmatrix}.
\end{split}
\end{equation}
For instance, for the bilayer model on a square lattice with only the in-plane nearest-neighbor hopping $t$ and the vertical inter-plane hopping $t_\perp$ [Fig.~18(a)], $\varepsilon_1(\bm{k})=\varepsilon_2(\bm{k})=2t[\cos(k_x)+\cos(k_y)]$, and $\varepsilon'(\bm{k})=t_\perp$.
With $R$, $\hat{H}(\bm{k})$ is transformed as

\begin{equation}
\begin{split}
&R\hat{H}(\bm{k})R^{\dagger}\\
&=\frac{1}{2}\begin{pmatrix} 
\varepsilon_{1}(\bm{k})+\varepsilon_{2}(\bm{k})+2\varepsilon'(\bm{k}) & \varepsilon_{2}(\bm{k})-\varepsilon_{1}(\bm{k}) \\ \varepsilon_{2}(\bm{k})-\varepsilon_{1}(\bm{k}) & \varepsilon_{1}(\bm{k})+\varepsilon_{2}(\bm{k})-2\varepsilon'(\bm{k}) \end{pmatrix}.
\end{split}
\end{equation}
We can thus see that the term $t_\perp$ contained in $\varepsilon'$ corresponds to the energy offset $\Delta E$ between the two orbitals.\\

\end{document}